\documentclass[a4paper,UKenglish,cleveref, autoref, thm-restate]{lipics-v2021}

\title{Shortest Beer Path Queries based on Graph Decomposition} 

\author{Tesshu Hanaka
}{Faculty of Information Science and Electrical Engineering, Kyushu University, Japan}{hanaka@inf.kyushu-u.ac.jp}{https://orcid.org/0000-0001-6943-856X}{JSPS KAKENHI Grant Numbers 
JP21H05852, 
JP21K17707, 
JP22H00513, 
and JP23H04388. 
}

\author{Hirotaka Ono
}{Graduate School of Informatics, Nagoya University, Japan 
}{ono@nagoya-u.jp}{https://orcid.org/0000-0003-0845-3947}{JSPS KAKENHI Grant Numbers JP20H00081, JP20H05967, JP21K19765, JP22H00513}

\author{Kunihiko Sadakane}{
Graduate School of Information Science and Technology, The University of Tokyo, Japan}{sada@mist.i.u-tokyo.ac.jp}{https://orcid.org/0000-0002-8212-3682}{JSPS KAKENHI Grant Numbers JP20H05967.}

\author{Kosuke Sugiyama 
}{Graduate School of Informatics, Nagoya University, Japan}{johnqpublic@dummyuni.org}{}{}

\authorrunning{Hanaka, et al. } 

\Copyright{Jane Open Access and Joan R. Public} 



\category{} 

\relatedversion{} 

\nolinenumbers 

\EventEditors{John Q. Open and Joan R. Access}
\EventNoEds{2}
\EventLongTitle{42nd Conference on Very Important Topics  (CVIT 2016 )}
\EventShortTitle{CVIT 2016}
\EventAcronym{CVIT}
\EventYear{2016}
\EventDate{December 24--27, 2016}
\EventLocation{Little Whinging, United Kingdom}
\EventLogo{}
\SeriesVolume{42}
\ArticleNo{23}

\newcommand{\ksplit}[1]{\mathrm{Spl}_{#1}}
\newcommand{\ksplitCom}[2]{\mathrm{SplCom}_{#1}  (#2 )}
\newcommand{\ksxy}[2]{\sigma^{\mathrm{xy}}_{#1}[{#2}]}
\newcommand{\ksyx}[2]{\sigma^{\mathrm{yx}}_{#1}[{#2}]}
\newcommand{\kbxx}[2]{\beta^{\mathrm{xx}}_{#1}[{#2}]}
\newcommand{\kbxy}[2]{\beta^{\mathrm{xy}}_{#1}[{#2}]}
\newcommand{\kbyx}[2]{\beta^{\mathrm{yx}}_{#1}[{#2}]}
\newcommand{\kbyy}[2]{\beta^{\mathrm{yy}}_{#1}[{#2}]}

\usepackage{diagbox} 
\usepackage{mathtools} 
\usepackage{multirow}

\newcommand{\kct}{\mathcal{T}} 
\newcommand{\kce}{\mathcal{E}}

\newcommand{\kskl}[1]{\mathrm{Sk}_{#1}} 
\newcommand{\kch}[1]{\mathrm{Ch}_{#1}} 
\newcommand{\kdes}[1]{\mathrm{Des}_{#1}} 

\newcommand{\kck}{\mathcal{K}} 
\newcommand{\krefe}[1]{\mathrm{Ref}_{#1}}
\newcommand{\kAlgD}[1]{\mathrm{ALG}\left ({#1}\right )}

\newcommand{\kdis}[3]{\mathrm{d}\!\left ({#1},{#2},{#3}\right )}
\newcommand{\kbdis}[3]{\mathrm{d^B}\!\left ({#1},{#2},{#3}\right )}

\newcommand{\kdsD}[4]{f_{#1}\!\left ({#2},{#3},{#4}\right )}
\newcommand{\kdsBD}[4]{f_{#1}^\mathrm{B}\!\left ({#2},{#3},{#4}\right )}
\newcommand{\kdsVec}[4]{\vec{f_{#1}}\!\left ({#2},{#3},{#4}\right )}
\begin{document}

\maketitle

\begin{abstract}
Given a directed edge-weighted graph $G=(V, E)$ with beer vertices $B\subseteq V$, a beer path between two vertices $u$ and $v$  is a path between $u$ and $v$ that visits at least one beer vertex in $B$, and the beer distance between two vertices is the shortest length of beer paths. 
We consider \emph{indexing problems} on beer paths, that is, a graph is given a priori, and we construct
some data structures (called indexes) for the graph. Then later, we are given two vertices, and we find the beer distance or
beer path between them using the data structure.
For such a scheme, efficient algorithms using indexes for the beer distance and beer path queries have been proposed for
outerplanar graphs and interval graphs. 
For example, Bacic et al. (2021) present indexes with size $O(n)$ for outerplanar graphs and an algorithm using them that answers the beer distance between given two vertices in $O(\alpha(n))$ time, where $\alpha(\cdot)$ is the inverse Ackermann function; the performance is shown to be optimal. 
This paper proposes indexing 
data structures 
and algorithms for beer path queries on general graphs based on two types of graph decomposition:
the tree decomposition and the triconnected component decomposition. 
We propose indexes with size $O(m+nr^2)$ based on the triconnected component decomposition, where $r$ is the size of the largest triconnected component. For a given query $u,v\in V$, our algorithm using the indexes can output the beer distance in query time $O(\alpha(m))$. 
In particular, 
our indexing data structures and algorithms achieve the optimal performance (the space and the query time) for series-parallel graphs, which is a wider class of outerplanar graphs.     
\end{abstract}

\section{Introduction}
Given a directed edge-weighted graph $G=(V, E)$ with beer vertices $B\subseteq V$, a beer path between two vertices $u$ and $v$  is a path between $u$ and $v$ that visits at least one beer vertex in $B$, and the beer distance between two vertices is the shortest length of beer paths. Here, a graph with $B$, the set of beer stores, is called a beer graph.
The names ``beer path'' and ``beer distance'' come from the following story: A person will visit a friend but does not want to show up empty-handed, and they decide to pick up some beer along the way. They would like to take the fastest way to go from their place to their friend's place while stopping at a beer store to buy some drinks. 

The notion of the beer path was recently introduced by Bacic et al.~\cite{bacic2023shortest}. 
Although the name is somewhat like a fable, we often encounter similar situations as the above story. Instead of beer stores, we want to stop at a gas station along the way, for example.  

Just computing the beer distance or a beer path with the beer distance is easy. A beer path with the beer distance always consists of two shortest paths: from 
the source to 
one of the beer stores
and from the beer store to the destination. 
We can 
therefore 
compute them by solving the single source shortest path problem twice 
from the source and from the destination, 
and taking the minimum beer vertex among $B$. 

We consider \emph{indexing problems} on beer paths, that is, a graph is given a priori, and we construct
some data structures (called indexes) for the graph. Then later, we are given two vertices, and we find the beer distance or
beer path between them using the data structure. This is more efficient than algorithms without using any indexes
if we need to solve queries for many pairs of vertices. 
Indeed, car navigation systems might equip such indexing mechanisms; since a system has map information as a graph in advance, it can make indexed information by preprocessing, which enables it to quickly output candidates of reasonable routes from the current position as soon as receiving a goal point. Such a scenario is helpful also for the beer path setting. 
Efficient algorithms using indexes for the beer distance and beer path queries have been proposed for
outerplanar graphs~\cite{bacic2023shortest} and interval graphs~\cite{das_et_al:LIPIcs.ISAAC.2022.59}.

This paper presents indexes with efficient query algorithms using graph decomposition for more general classes of graphs. 
Namely, we consider graphs of bounded treewidth~\cite{bodlaender1997treewidth}
and graphs with bounded triconnected components size~\cite{doi:10.1137/0202012}.
The performance of our indexing and algorithm generalizes that for outerplanar graphs in \cite{bacic2023shortest}; 
if we apply our indexing and algorithm for outerplanar graphs, the space for indexes, preprocessing time, and query time are equivalent to 
those of \cite{bacic2023shortest}. Furthermore, ours can be applied for general graphs, though the performance worsens for graphs of large treewidth and with a large triconnected component.  

\subsection{Related Work}
For undirected outerplanar beer graphs $G$ of order $n$, Bacic et al.~\cite{bacic2023shortest} present indexes with size $O(n)$, which can be preprocessed in $O(n)$ time. For any two query vertices $u$ and $v$, (i) the beer distance between $u$ and $v$ can be reported in $O(\alpha(n))$ time, where $\alpha(n)$ is the inverse Ackermann function, and (ii) a beer path with the beer distance between $u$ and $v$ can be reported in $O(L)$ time, where $L$ is the number of vertices on this path. The query time is shown to be optimal. 

For unweighted interval graphs with beer vertices $B$, Das et al.~\cite{das_et_al:LIPIcs.ISAAC.2022.59} provides a representation using $2n \log n + O(n) + O(|B| \log n)$ bits. This data structure answers beer distance queries in $O(\log^\varepsilon n)$ time for any constant $\varepsilon > 0$ and shortest beer path queries in $O(\log^\varepsilon n + L)$ time. They also present a trade-off relation between space and query time. These results are summarized in Table~\ref{tab:genresults}.
\begin{table}[hbtp]
    \centering
    \caption{Comparison of results}\label{tab:genresults}
    \begin{tabular}{cc||c|c|c}
        \hline 
      \multicolumn{2}{c||}{\multirow{2}{*}{Graphs}} & \multicolumn{2}{c|}{Preprocessing Complexity} & \multirow{2}{*}{Query time}\\ \cline{3-4}
      &   &  Space (words) & Time & \\ \hline\hline 
      \multicolumn{2}{c||}{Outerplanar graphs~\cite{bacic2023shortest}} &  \multicolumn{2}{c|}{\multirow{2}{*}{$O(m)(=O(n))$}}  & \multirow{2}{*}{$O(\alpha(n))$} \\ 
     \multicolumn{2}{c||}{(undirected, $W=\mathbb{R}_{\geq 0}$)} &  \multicolumn{2}{c|}{}  &  \\ \hline 
      \multicolumn{2}{c||}{Interval graphs~\cite{das_et_al:LIPIcs.ISAAC.2022.59}}& {$O(n+|B|)$} & {$O(n+|B|)$}$^*$ & $O(\log^{\varepsilon} n)$ \\ 
       \multicolumn{2}{c||}{(undirected, unweighted)} & {$O(n+|B|\log\log n)$} & {$O(n+|B|\log\log n)$}$^*$ & $O(\log\log n)$ \\ \hline       
       \multirow{6}{*}{Ours} & tri. decomp. &     $O(m)$ & $O(m+nr^3)$ & $O(r^2+\alpha(m))$ \\ 
                             & (undirected, $W=\mathbb{Z}_{\geq 0}$) & $O(m+nr^2)$ & $O(m+nr^4)$ & $O(\alpha(m))$ \\ \cline{2-5} 
                             & tri. decomp. &$O(m)$ & $O((m+n\log r_+)r_+^2)$ & $O(r^2\log r_+ +\alpha(m))$ \\ 
                             &  ({$W=\mathbb{R}_{\geq 0}$}) & $O(m+nr^2)$ & $O((m+n\log r_+)r_+^3)$ & $O(\alpha(m))$ \\ \cline{2-5}
       & tree decomp. & $O(t^3 n)$ & $O(t^8 n)$ & $O(t^7 +\alpha(tn))$ \\ 
       & ($W=\mathbb{R}_{\geq 0}$) & $O(t^5 n)$ & $O(t^{10} n)$ & O($t^6 + \alpha(tn))$ \\ \hline 
    \end{tabular}
    \begin{quote}
    $n$: the number of vertices \\ 
    $m$: the number of edges \\ 
    $r$: the size of maximum triconnected components, $r_+=\max (1,r)$ \\
    $t$: the treewidth \\ 
    $\alpha(\cdot)$: the inverse Ackermann function \\ 
    $^*$: not explicitly mentioned \\
    \end{quote}
    
\end{table}

Other than the beer path problem, Farzan and Kamali~\cite{DBLP:journals/algorithmica/FarzanK14} proposed a distance oracle for graphs with $n$ vertices
and treewidth $k$ using asymptotically optimal $k(n+o(n)-k/2)+O(n)$ bit space which can answer a query in $O(k^3 \log^3 k)$ time.
%
For general graphs, indexes for shortest paths (i.e., distance queries) and max flow queries 
based on the triconnected component decomposition have been proposed~\cite{DBLP:journals/jgaa/KashyopNS19}.

\subsection{Our Contribution}
We present indexing data structures and query algorithms for beer distance and beer path queries for general graphs based on graph decomposition.
As graph decomposition, we use the tree decomposition~\cite{bodlaender1997treewidth} and the triconnected component decomposition~\cite{doi:10.1137/0202012}.
The obtained results are summarized in Table \ref{tab:genresults}. 

We first present faster query algorithms using properties of the triconnected component decomposition.
In this approach, we use $r$, the size of the largest triconnected component in a graph, as the parameter to evaluate the efficiency
of algorithms.  Note that $r$ is not the number of edges in the largest triconnected component; it is the
number of edges in a component after contracting every biconnected component into an edge and therefore it is not so large in practice.
The formal definition will be given in Section~\ref{sec:SPQR}.
Our data structure uses $O(m+r\cdot\min\{m,rn\})$ space, and the algorithm for undirected graphs with nonnegative integer edge weights requires $O(m+r^3\cdot\min\{m,rn\})$ time for preprocessing, and it answers for each query in $O(\alpha(m))$ time.   
For directed graphs with nonnegative edge weights, the preprocessing time and query time are, respectively 
$O(m+r^3(m+n\log r_+))$ and $O(\alpha(m))$. Since the size of indexes and query time have a trade-off relation, 
a little slower query time can achieve an indexing data structure with less memory. In such a scenario, 
another data structure uses $O(m)$ space, and the algorithm for undirected graphs with nonnegative integer edge weights requires $O(m+r^2\cdot\min\{m,rn\})$ time for preprocessing, and it answers for each query in $O(r^2+\alpha(m))$ time.   
For directed graphs with nonnegative edge weights, the preprocessing time and query time are, respectively 
$O(m+r^2(m+n\log r_+))$ and $O(r^2\log r_+ +\alpha(m))$.

Because triconnected component decomposition can be regarded as a tree decomposition, we extend our query algorithms for
graphs represented by using the tree decomposition.
Though computing the exact treewidth is NP-hard, whereas
triconnected component decomposition is done in linear time~\cite{10.1007/3-540-44541-2_8},
and query time complexities using tree decomposition is larger than using triconnected component decomposition, 
the treewidth is always at most $r$ and therefore algorithms based on the tree decomposition are
faster in some cases.
In view of these, we remake the indexing data structures and algorithms for tree decomposition. 
The indexing data structure requires $O(t^5n)$ space and $O(t^{10}n)$ time to construct, and the algorithm can answer a query in $O(t^6+\alpha(tn))$ time. 

Note that for series-parallel graphs $r=0$, $t=2$, and $m=O(n)$ hold. This implies that for series-parallel graphs, our indexing data structures use $O(n)$ space, and the algorithms can answer each query in $O(\alpha(n))$ time. Since the class of series-parallel graphs is a super class of outerplanar graphs, our results fairly extend the optimal result for outerplanar graphs by \cite{bacic2023shortest}. 

\medskip 

The rest of the paper is organized as follows. 
Section~\ref{sec:preliminaries} is for preliminaries. 
Sections~\ref{sec:triindex} and~\ref{sec:trialgo} present the main parts that describe the indexing and algorithms under triconnected decomposition. 
Section~\ref{sec:treedecomposition} shows how we remake that to those under tree decomposition. 

\section{Preliminaries}\label{sec:preliminaries}
Let $\mathbb{Z}_{\geq 0}$ be the set of nonnegative integers and $\mathbb{R}_{\geq 0}$ be the set of nonnegative real numbers. 
For nonnegative integers $i,j \in \mathbb{Z}_{\geq 0}$  ($i\leq j$), let $[i,j]=\{i,i+1,\ldots, j-1,j\}$.

For a graph $G$, let $V (G)$ and $E(G)$ denote its vertex and edge sets, respectively. For two graphs $G$ and $G'$, let $G\setminus G'=\left (V (G )\setminus V (G' ),E (G )\setminus E (G' )\right )$ and $G\cup G'=\left (V (G )\cup V (G' ),E (G )\cup E (G' )\right )$. Also, for a graph $G$ and a set of vertex pairs $F \subseteq V (G )\times V (G )$, let $G\setminus F =  (V (G ),E (G )\setminus F )$ and $G\cup F =  (V (G ),E (G )\cup F )$. Furthermore, for a graph $G$ and its vertex subset $S\subseteq V (G )$, let $G[S]$ be the subgraph of $G$ induced by $S$.

\subsection{Shortest Path Problem and Beer Path Problem / Query}
Suppose we are given a graph $G$, an edge weight $w\colon E (G ) \to W$, and a vertex subset $B\subseteq V (G )$. Note that in this paper, we assume $W=\mathbb{Z}_{\geq 0}$ or $W=\mathbb{R}_{\geq 0}$.

For vertices $u,v\in V (G )$, a path from $u$ to $v$ in $G$ is called a \emph{$u$-$v$ path} in $G$. Usually, a $u$-$v$ path is not unique.
The length of a path is defined by the sum of the edge weights on the path. The shortest length of all $u$-$v$ paths is called the $u$-$v$ distance in $G$, denoted by $\kdis{ (G,w)}{u}{v}$. 
Also, for vertices $u,v\in V (G )$, a walk from $u$ to $v$ passing through a vertex belonging to $B$ at least once is called a 
\emph{$u$-$v$ beer path} in $G$. The length of a $u$-$v$ beer path is similarly defined as the length of a $u$-$v$ path, 
$u$-$v$ beer distance in $G$ is defined by the shortest length of all $u$-$v$ beer paths and is denoted by $\kbdis{ (G,w,B )}{u}{v}$.  
Note that if $w$ and $B$ are clear from the context, we omit them and denote $\kdis{ (G,w )}{u}{v}$ as $\kdis{G}{u}{v}$ and $\kbdis{ (G,w,B )}{u}{v}$ as $\kbdis{G}{u}{v}$. Then, a vector whose elements are the distance and the beer distance is denoted by 
\[
\vec{d} (G,u,v ) = \begin{pmatrix} \kdis{G}{u}{v} \\ \kbdis{G}{u}{v} \end{pmatrix}. 
\]
For a given $G,w,B$ and vertices $u,v\in V (G )$, the problem of finding $\kdis{G}{u}{v}$  (or one $u$-$v$ path that realizes it) is called \textsc{Shortest Path} and the problem of finding $\kbdis{G}{u}{v}$  (or one $u$-$v$ beer path that realizes it) is called \textsc{Beer Path}. 
For given $u$ and $v$, the query asked to return $\kbdis{G}{u}{v}$ or one $u$-$v$ beer path with length $\kbdis{G}{u}{v}$ 
is called \emph{Beer Path Query} on $G,w,B$.

Here, we review algorithms for the shortest path problem and their computational complexity. When $W=\mathbb{Z}_{\geq 0}$ and $G$ is an undirected graph, the shortest path problem can be solved in $O(m)$ time by using Thorup's algorithm~\cite{thorup1999undirected}. When $W=\mathbb{R}_{\geq 0}$, the shortest path problem can be solved in $O(m+ n\log n)$ time by Dijkstra's algorithm using Fibonacci heap~\cite{fredman1987fibonacci}. Hereafter, 
let $\kAlgD{G}$ denote the computational time to solve \textsc{Shortest Path Problem} by one of the above algorithms 
according to the setting; for example, if $G$ is undirected and $W=\mathbb{Z}_{\geq 0}$, $\kAlgD{G}=O(m)$, and if $W=\mathbb{R}_{\geq 0}$, 
$\kAlgD{G}=O(m+n\log m)$. 

\subsection{SPQR tree}\label{sec:SPQR}
Let $G$ be a biconnected (multi) undirected graph and $\{u,v\}$ be its vertex pair. 
If $G[V(G)\setminus\{u,v\}]$ is 
disconnected 
or $u$ and $v$ are adjacent in $G$, $\{u,v\}$ is called a split pair of $G$. 
We denote the set of split pairs of $G$ by $\ksplit{G}$. 
For $\{u,v\} \in \ksplit{G}$, a maximal subgraph $H$ of $G$ satisfying $\{u,v\} \notin \ksplit{H}$, 
and the graph $(\{u,v\},\{e\})$ consisting of the edge $e\in E(G)$, 
are called a split component of the split pair $\{u,v\}$ of $G$. 
We denote the set of split components of the split pair $\{u,v\}$ of $G$ by $\ksplitCom{G}{u,v}$. 
For $\{u,v\}\in \ksplit{G}$ and $\{s,t\}\in E(G)$, 
we say that $\{u,v\}$ is maximal 
with respect to $\{s,t\}$ 
if vertices $u,v,s,t$ are in the same split component.

For an edge $e=\{u,v\} \in E(G)$, we define an SPQR tree $\kct(G,e)$ of $G$. Here, $e$ is called a \emph{reference edge} of $\kct(G,e)$ of $G$. 
Each node $\mu$ of $\kct(G,e)$ is associated with a graph $\kskl{\mu}$. The root node of $\kct(G,e)$ is denoted by $\mu_e$. The $\kct(G,e)$ is defined recursively as follows.
\begin{description}
    \item[Trivial Case] If $\ksplitCom{G}{u,v}=\{(\{u,v\},\{e\}),(\{u,v\},\{e'\})\}$ ($e'\in E(G)$), that is, $G$ is a two vertices multi graph consisting of two edges $e,e'$, $\kct(G,e)=(\{\mu_e\}, \emptyset)$, $\kskl{\mu_e}=G$. Also, $\mu_e$ is said to be a Q node.
    \item[Series Case] Let $\ksplitCom{G}{u,v}=\{(\{u,v\},\{e\}),H\}$, where $H$ is formed by a series connection of $k (\geq 2)$ connected components $H_1,\ldots H_k$. Then, for vertices $u=c_0, c_1, c_2,\ldots ,c_{k-1},$ $c_k =v$ ($c_1, \ldots , c_{k-1}$ are cut vertices of $G$), let $c_{i-1},c_i$ be the only vertices belonging to $H_i$ ($1\leq i\leq k$). In this case, if $e_i=\{c_{i-1},c_i\}$ ($1\leq i\leq k$), then
    $$\kct(G,e)=(\{\mu_e\},\emptyset)\cup \bigcup_{1\leq i\leq k}\left( \kct(H_i\cup \{e_i\},e_i) \cup \{\{\mu_e,\mu_{e_i}\}\}\right),$$
    $$\kskl{\mu_e}=(\{c_0,\ldots,c_k\},\{e,e_1,\ldots,e_k\}).$$
    Also, $\mu_e$ is said to be an S node.
    \item[Parallel Case] If $\ksplitCom{G}{u,v}=\{(\{u,v\},\{e\}),H_1,\ldots,H_k\}$ ($k\geq 2$), that is,  $G$ is formed by the parallel connection of 3 or more split components of $\{u,v\}$. In this case, if we let $e_i$ denote the edge corresponding to $H_i$, then $\kskl{\mu_e}=(\{u,v\},\{e,e_1,\ldots,e_k\})$ and $\kct(G,e)$ is defined as same as series case. Also, $\mu_e$ is said to be a P node.
    \item[Rigid Case] If the above does not apply, that is, $\ksplitCom{G}{u,v}=\{(\{u,v\},\{e\}),H\}$ and $H$ has no cut vertices, let all maximal split pairs for $\{u,v\}$ in $\ksplit{G}\setminus \{\{u,v\}\}$ be $\{u_i,v_i\}$ ($1\leq i\leq k, k\geq 1$). Also, for each $i$, let $H_i$ be the union of the split components for $\{u_i,v_i\}$ that does not contain $e$. That is, $H_i = \bigcup_{H\in \ksplitCom{G}{u_i,v_i}:e\notin E(H)} H$. In this case, if we let $e_i$ denote the edge corresponding to $H_i$ ($1\leq i\leq k$), then
    $$\kskl{\mu_e}=\left(\{u,v\}\cup\bigcup_{1\leq i\leq k} \{u_i,v_i\}, \{e\}\cup\bigcup_{1\leq i\leq k} \{e_i\} \right)$$
    and $\kct(G,e)$ is defined as same as series case. Also, $\mu_e$ is said to be an R node.
\end{description}
The tree $(\{\rho,\emptyset\})\cup \kct(G,e) \cup \{\{\rho,\mu_e\}\}$ obtained by connecting the tree $\kct(G,e)$ obtained by the above definition and Q node $\rho$ with the graph $\kskl{\rho}=(\{u,v\},\{e\})$ as a root is called the SPQR tree of $G$ with respect to edge $e$. Hereafter, we simply call it an SPQR tree and denote it by $\kct$. Also, we denote the only child node $\mu_e$ of the root node $\rho$ by $\rho'$.

\begin{figure}[h]
    \centering
    \includegraphics[keepaspectratio, scale=0.5]{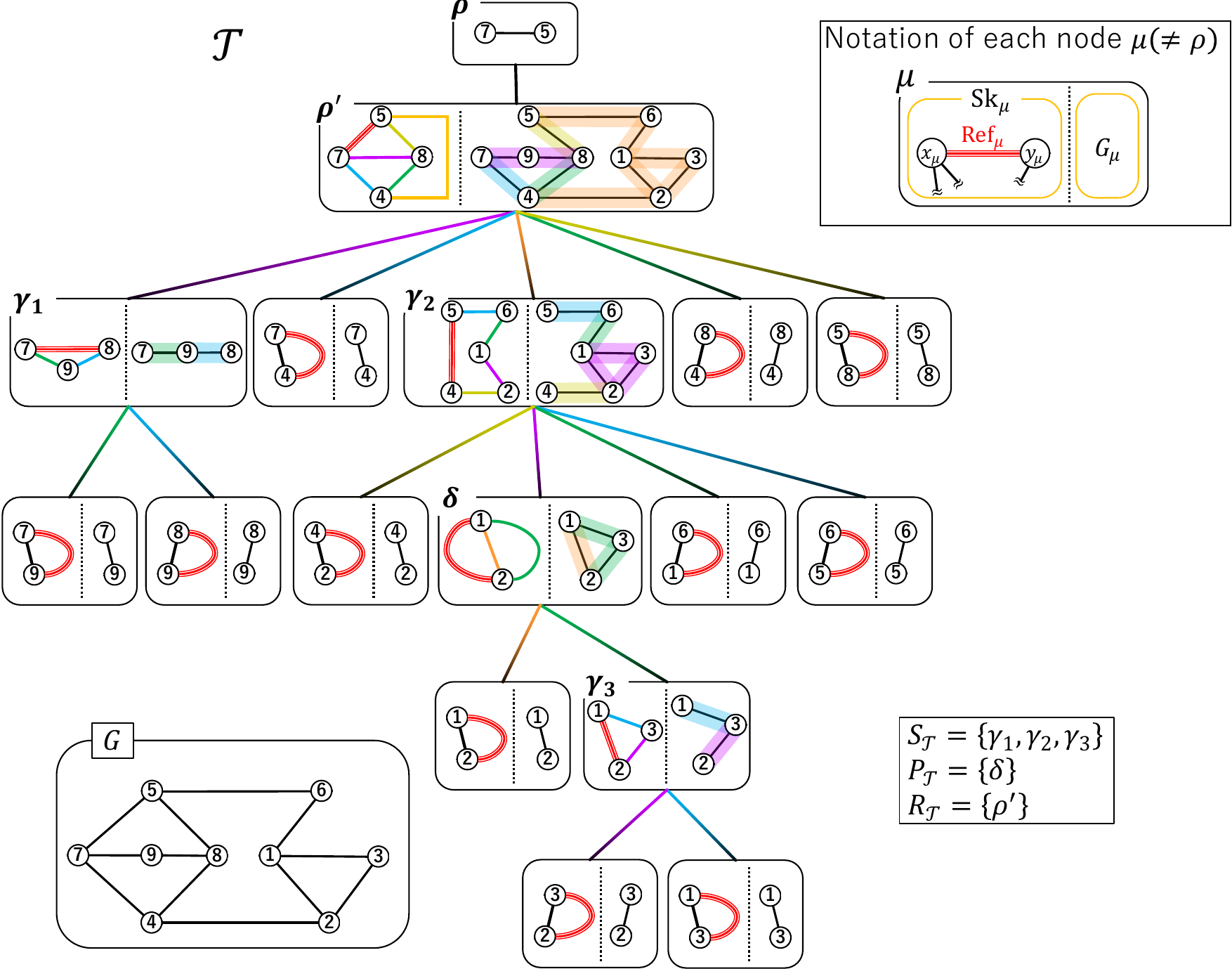}
    \caption{An graph $G$ (lower left) and its SPQR tree $\kct$}
    \label{fig:SPQRtree}
\end{figure}

For each node $\mu \in V(\kct)\setminus \{\rho\}$ of $\kct$, each edge of $\kskl{\mu}$ is a skeleton of a certain graph, so $\kskl{\mu}$ is called the skeleton graph of $\mu$. Let $n_{\mu} = \left| V(\kskl{\mu}) \right|$ be the number of vertices and $m_{\mu} = \left| E(\kskl{\mu}) \right|$ be  the number of edges of the skeleton $\kskl{\mu}$. Also, let the reference edge of $\mu$ be $\krefe{\mu}=\{x_{\mu}, y_{\mu}\}$ and let $\kch{\mu}$ and $\kdes{\mu}$ be the sets of child and descendant nodes of $\mu$ in $\kct$, respectively. We denote the set consisting of S, P, Q, and R nodes by $S_{\kct}, P_{\kct}, Q_{\kct}, R_{\kct}$, respectively.
For each $\mu \in V(\kct)\setminus \{\rho\}$, let $G_{\mu}$ be the subgraph of $G$ corresponding to the graph of $\kskl{\mu}$ without the reference edge. This can be expressed as $G_{\mu}=(\{x_{\mu}, y_{\mu}\},\{\{x_{\mu}, y_{\mu}\}\})$ if $\mu\in Q_\kct$, 
otherwise $G_\mu=\bigcup_{\lambda\in\kch{\mu}} G_\lambda$. An example of a SPQR tree is shown in Figure~\ref{fig:SPQRtree} of Apendix.

The following is known for the SPQR tree $\kct$ of $G$. 
\begin{lemma}
    Let $G$ be a biconnected undirected graph with $n$ vertices and $m$ edges, and $\kct$ be its SPQR tree. For each node $\mu\in V(\kct)\setminus\{\rho\}$, $\{x_{\mu}, y_{\mu}\}\in \ksplit{G}$. If $\mu\in R_\kct$, $\kskl{\mu}$ is a triconnected graph. Also, $|Q_\kct|=m$, $|S_\kct \cup P_\kct \cup R_\kct| = O(n)$, $\sum_{\mu \in S_\kct \cup P_\kct \cup R_\kct}m_{\mu}=O(m)$, and $\sum_{\mu \in V(\kct)}n_\mu = O(n)$ hold. Furthermore, $\kct$ van be computed in $O(n+m)$ time.
\end{lemma}

Let $r = \max_{\mu \in R_\kct} \{ m_\mu \}$ be the maximum number of edges in the skeleton of the R node (triconnected graph). Note that if $R_\kct = \emptyset$), $r=0$. Also, $r_{+} = \max \{1,r \}$.

An SPQR tree for a directed graph is defined as a graph whose skeleton is replaced by a directed graph after computing the SPQR tree by considering the graph as an undirected graph. In this case, for each $\mu\in V(\kct)\setminus\{\rho\}$, we consider two reference edges $\langle x_\mu,y_\mu \rangle,\langle y_\mu,x_\mu \rangle$ and let $\krefe{\mu}=\{\langle x _\mu,y_\mu \rangle, \langle y_\mu,x_\mu \rangle\}$.


\subsection{Query Problems}
We present all query problems that will be used in later.
\begin{description}
    \item[Range Minimum Query] For a given array $(a)=a[1],a[2],\ldots,a[n]$ of length $n$, the query defined by the following pair of inputs and outputs is called Range Minimum Query for the array $(a)$.
    \begin{description}
        \item[Input] Positive integers $i,j$ ($1\leq i\leq j \leq n$),
        \item[Output] Minimum value in the subarray $a[i],a[i+1],\ldots,a[j-1],a[j]$ of the array $(a)$.
    \end{description}
\end{description}
This query can be answered in $O(1)$ time by preprocessing in $O(n)$ space and $O(n)$ time~\cite{doi:10.1137/0222017}.

\begin{description}
    \item[Lowest Common Ancestor Query] Given a rooted tree $T$ with $n$ vertices. The query defined by the following pair of input and output is called the lowest common ancestor query for the rooted tree $T$.
    \begin{description}
        \item[Input] Vertices $u,u'\in V(T)$,
        \item[Output] The deepest (furthest from the root) common ancestor of $u,u'$ in $T$.
    \end{description}
\end{description}
This query can be answered in $O(1)$ time by preprocessing in $O(n)$ space and $O(n)$ time using range minimum queries.

\begin{description}
    \item[Tree Product Query] Given a set $S$, a semigroup $\circ\colon S^2 \to S$, a tree $T$ with $n$ vertices, and a mapping $f\colon V(T) \to S$. The query defined by the following pair of input and output is called a Tree Product Query for $S,\circ,T,f$.
    \begin{description}
        \item[Input] Vertices $u,u'\in V(T)$,
        \item[Output] Let $u=v_1, v_2,\ldots,v_{k-1},v_k=u'$ be the only path on $T$ that connects $u,u'$, then $f(v_1)\circ f(v_2)\circ \ldots \circ f(v_k)$.
    \end{description}
\end{description}
This query can be answered in $O(\alpha(n))$ time by preprocessing in $O(n)$ space and $O(n)$ time~\cite{1360011146378384384}. 
Here, $\alpha$ is the inverse Ackermann function.

For a nonnegative integer $i,\ell$, we define $A_{\ell}(i)$ as follows.
$$A_{\ell}(i)=\left\{\begin{matrix}
    i+1 & \ell = 0, i\geq 0\\
    A_{\ell -1}^{(i+1)}(i+8) & \ell \geq1,i \geq 0
\end{matrix}\right. .$$
Note that $A_{\ell -1}^{(i+1)}$ denotes a function that $A_{\ell -1}$ iterated $i+1$ times. Using this, the inverse Ackermann function $\alpha$ is defined as follows.
$$\alpha(n) = \min \{ \ell \in \mathbb{Z}_{\geq 0} \mid A_{\ell}(1)>n \}.$$
\section{Triconnected component decomposition-based indexing}\label{sec:triindex}
This section describes indexes based on triconnected component decomposition for biconnected graphs. First, for each $\mu,\lambda \in \kct$, we define $K_{\mu,\lambda} = \left ( \{x_{\mu},y_{\mu}\} \cup \{x_{\lambda},y_{\lambda}\} ,  (\{x_{\mu},y_{\mu}\} \cup \{x_{\lambda},y_{\lambda}\} )^2 \right )$ to be a complete graph with self loops, and let $K_{\mu,\lambda}^{\vec{w}}$ denote the graph $K_{\mu,\lambda}$ with the weight
$$\vec{w}\colon V (K_{\mu,\lambda} )^2 \to W^2 \ (\vec{w} (u,v )=\begin{pmatrix} w (u,v )\\w^\mathrm{B} (u,v )\end{pmatrix}).$$
Also, let
$$\kck = \left\{ K_{\mu,\lambda}^{\vec{w}} \mid  \mu,\lambda \in V (\kct ), \vec{w}\colon V (K_{\mu,\lambda} )^2 \to W^2 \right\} \cup \{\perp\}$$
be the union of the set of those weighted graphs and $\{\perp\}$.

Next, for convenience, we define the maps $F_i \colon \mathrm{dom} (F_i )\to \kck$ that provide data of distance and beer distance  ($i=1,2,3,4$, specific domains are described later). The algorithms for beer path queries precompute some of these as data structures.

For each $\mathcal{X} \in \mathrm{dom} (F_i )$, let $F_i (\mathcal{X} )\in \kck$ be a complete graph with at most 4 vertices and $\vec{f}_i (\mathcal{X} )$ be its weight. Also, we will denote the weight $\vec{f}_i (\mathcal{X} ) (u,v )$ of each vertex pair $\langle u,v \rangle \in V (F_i (\mathcal{X} ) )^2$ by
$$\kdsVec{i}{\mathcal{X}}{u}{v}=\begin{pmatrix} \kdsD{i}{\mathcal{X}}{u}{v} \\ \kdsBD{i}{\mathcal{X}}{u}{v}\end{pmatrix},$$
omitting some brackets. These maps are defined so that $\kdsD{i}{\mathcal{X}}{u}{v}$ represents the normal distance and $\kdsBD{i}{\mathcal{X}}{u}{v}$ represents the beer distance.
\subsection{Definition of the mapping $F_1$ and its computation}
We define the mapping $F_1 \colon V (\kct )\setminus \{\rho\} \to \kck$ as follows.
\begin{definition}\label{def:mapF1}
    For each node $\mu \in V (\kct )\setminus \{\rho\}$, let $F_1 (\mu )=K_{\mu,\mu}^{\vec{f}_1 (\mu )}$ (a complete graph consists of 2 vertices $x_{\mu}, y_{\mu}$). The weight of each vertex pair $\langle u,v \rangle \in \{x_{\mu}, y_{\mu}\}^2$ is $\kdsVec{1}{\mu}{u}{v} = \vec{d} (G_{\mu}, u,v )$.
\end{definition}
The $F_1(\mu)$ intuitively represents the distance data when using the part of the $\kct$ shown in Figure~\ref{fig:SubgraphForF1} of Appendix.

We can compute $F_1$ from the leaves of $\kct$ to the root as described below.

If $\mu \in Q_{\kct}\setminus \{\rho \}$, $G_{\mu}$ is a graph that consists of only edges $\langle x_{\mu},y_{\mu}\rangle,\langle y_{\mu},x_{\mu}\rangle$, so each weight $\kdsVec{1}{\mu}{u}{v}$ can be calculated as in Table~\ref{table:F1_Qnode}.
From now on, we assume that $\mu$ is an inner node and that $F_1 (\lambda)$ is computed for each of its child nodes $\lambda \in \kch{\mu}$. Also, let $H_{\mu}$ be the weighted graph with each edge $\langle x_{\lambda},y_{\lambda} \rangle, \langle y_{\lambda},x_{\lambda} \rangle$  ($\lambda \in \kch{\mu}$) of $\kskl{\mu}\setminus \krefe{\mu}$ given a weight $\kdsD{1}{\mu}{x_\mu}{y_\mu},\kdsD{1}{\mu}{y_\mu}{x_\mu}$ respectively. Then, from the definition of $H_{\mu}$, if we consider the path from $u$ to $v$ in $G_{\mu}$ through the subgraph $G_{\lambda}$  ($\langle u,v \rangle \in \{ x_{\lambda},y_{\lambda} \}^2$).
The distance $\kdsD{1}{\mu}{u}{v}$ can be obtained by referring to the weight of the edge $\langle u,v \rangle \in E (H_{\mu} )$  (the beer distance $\kdsBD{1}{\mu}{u}{v}$ is obtained by referring to $\kdsBD{1}{\lambda}{u}{v }=\kbdis{G_{\lambda}}{u}{v}$ directly). Therefore, $F_1 (\mu )$ can be calculated by using $H_{\mu}$ instead of $G_{\mu}$.

If $\mu \in S_{\kct}$, let $\kch{\mu}=\{ \mu_1, \ldots , \mu_k \}$ and let $x_{\mu}=x_{\mu_1}, y_{\mu_{i}}=x_{\mu _{i+1}}\  (1 \leq i \leq k-1 ), y_{\mu_k}=y_{\mu}$ in $\kskl{\mu}$ (see Figure~\ref{fig:SnodeAndChildren} of Appendix). Here, we define the following six symbols for each $\mu\in S_\kct$.


\begin{itemize}
    \item $\ksxy{\mu}{i,j} \coloneqq \left\{ \begin{matrix}
        \sum_{i\leq p \leq j} \kdsD{1}{\mu_p}{x_{\mu_p}}{y_{\mu_p}} & 1\leq i \leq j \leq k, \\
        0 & \mathrm{otherwise,}
    \end{matrix} \right.$\\
    which is the distance from $x_{\mu_i}$ to $y_{\mu_j}$ in $\bigcup_{i\leq p\leq j}G_{\mu_p}$. 
    \item $\ksyx{\mu}{i,j} \coloneqq \left\{ \begin{matrix}
        \sum_{i\leq p \leq j} \kdsD{1}{\mu_p}{y_{\mu_p}}{x_{\mu_p}} & 1\leq i \leq j \leq k,\\
        0 & \mathrm{otherwise,}
    \end{matrix} \right.$\\
    which is the distance from $y_{\mu_j}$ to $x_{\mu_i}$ in $\bigcup_{i\leq p\leq j}G_{\mu_p}$. 
    \item $\kbxx{\mu}{i} \coloneqq \ksxy{\mu}{1,i-1} + \kdsBD{1}{\mu_i}{x_{\mu_i}}{x_{\mu_i}} + \ksyx{\mu}{1,i-1}$ ($1\leq i \leq k$, \\
    which is the distance of the shortest walk that reaches from $x_\mu=x_{\mu_1}$ to $y_{\mu_{i-1}}=x_{\mu_i}$ in $\bigcup_{1\leq j\leq i-1}G_{\mu_j}$, back to $x_{\mu_i}$ via a beer vertex in $G_{\mu_i}$ and again to $x_\mu$. 
    \item $\kbxy{\mu}{i} \coloneqq \kdsBD{1}{\mu_i}{x_{\mu_i}}{y_{\mu_i}}-\kdsD{1}{\mu_i}{x_{\mu_i}}{y_{\mu_i}}$ ($1\leq i \leq k$), \\
    which is the difference between the beer distance and the (mere) distance in moving from $x_{\mu_i}$ to $y_{\mu_i}$ in $G_{\mu_i}$. 
    \item $\kbyx{\mu}{i} \coloneqq \kdsBD{1}{\mu_j}{y_{\mu_j}}{x_{\mu_j}}-\kdsD{1}{\mu_j}{y_{\mu_j}}{x_{\mu_j}}$ ($1\leq i \leq k$), \\
    which is the difference between the beer distance and the (mere) distance in moving from $y_{\mu_i}$ to $x_{\mu_i}$ in $G_{\mu_i}$. 
    \item $\kbyy{\mu}{i} \coloneqq \ksyx{\mu}{i+1,k} + \kdsBD{1}{\mu_i}{y_{\mu_i}}{y_{\mu_i}} + \ksxy{\mu}{i+1,k}$ ($1\leq i \leq k$),\\
    which is the distance of the shortest walk that reaches from $y_\mu=y_{\mu_k}$ to $x_{\mu_{i+1}}=y_{\mu_i}$ in $\bigcup_{i+1\leq j\leq k}G_{\mu_j}$, back to $y_{\mu_i}$ via a beer vertex in $G_{\mu_i}$ and again to $y_\mu$.
\end{itemize}
Note that we only preprocess $\ksxy{\mu}{1,i},\ksxy{\mu}{i,k},\ksyx{\mu}{1,i},\ksyx{\mu}{i,k}$ ($1\leq i \leq k$) among $\ksxy{\mu}{i,j},\ksyx{\mu}{i,j}$. The other $\ksxy{\mu}{i,j}$ and $\ksyx{\mu}{i,j}$ are obtained and used in $O(1)$ time each time. We also preprocess $\kbxx{\mu}{\cdot},\kbxy{\mu}{\cdot},\kbyx{\mu}{\cdot},\kbyy{\mu}{\cdot}$ for Range Minimum Query. All of the above preprocessing can be computed in $O(k)$ space and $O(k)$ time.

By using these, each weight $\kdsVec{1}{\mu}{u}{v}$ can be calculated as in Table~\ref{table:F1_Snode}.

If $\mu \in P_{\kct}$, we define the following for each $\langle u,v \rangle \in \{x_{\mu}, y_{\mu}\}^2$ to simplify:
\[
\ell_{u,v}=\min_{\lambda \in \kch{\mu}} \left\{ \kdsD{1}{\lambda}{u}{v} \right\},\ \ell^\mathrm{B}_{u,v}=\min_{\lambda \in \kch{\mu}} \left\{ \kdsBD{1}{\lambda}{u}{v} \right\}.
\]


If $\mu \in R_{\kct}$, each weight $\kdsVec{1}{\mu}{u}{v}$ can be calculated on $H_{\mu}$ as follows.
$$\kdsVec{1}{\mu}{u}{v}
=\begin{pmatrix}
    \kdis{H_{\mu}}{u}{v}\\
    \min_{\lambda \in \kch{\mu}} \{ \min_{p,q \in \{x_{\lambda},y_{\lambda}\}} \{ \kdis{H_{\mu}}{u}{p} + \kdsBD{1}{\lambda}{p}{q} + \kdis{H_{\mu}}{q}{v} \}\}
\end{pmatrix}.
$$
Note that each $\kdis{H_{\mu}}{a}{b}$ in the above equation is calculated by a shortest path algorithm for $H_\mu$. An example of calculating $F_1$ is shown in Figure~\ref{fig:SPQRtree_F1F2} of Appendix.

\subsection{Definition of the mapping $F_2$ and its computation}
\begin{figure}[h]
    \centering
    \includegraphics[keepaspectratio, scale=0.5]{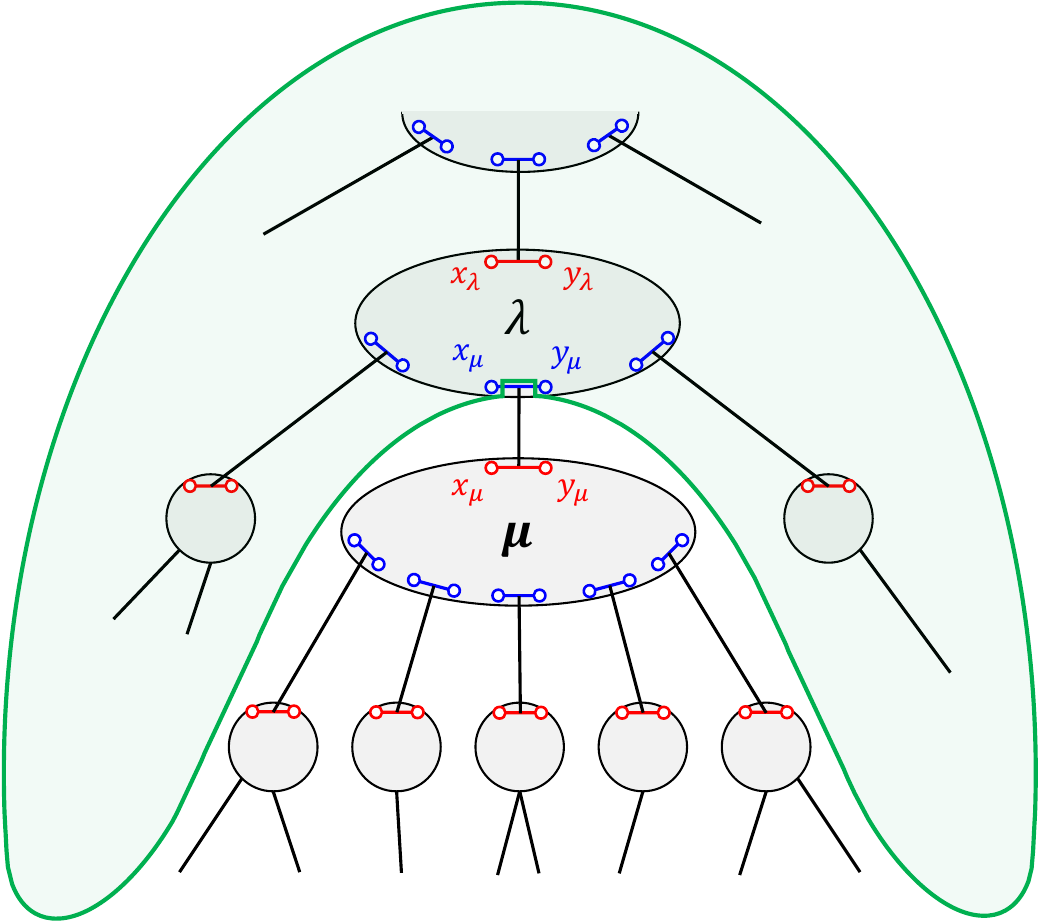}
    \caption{The subtree considered in  $F_2(\mu)$}
    \label{fig:SubgraphForF2}
\end{figure}

We define the mapping $F_2\colon V (\kct )\setminus \{\rho\} \to \kck$ as follows.
\begin{definition}\label{def:mapF2}
    For each node $\mu \in V (\kct )\setminus \{\rho\} $, let $F_2 (\mu )=K_{\mu}^{\vec{f_2} (\mu )}$ (a complete graph consists of 2 vertices $x_{\mu}, y_{\mu}$). The weight of each vertex pair $\langle u,v \rangle \in \{x_{\mu}, y_{\mu}\}^2$ is $\kdsVec{2}{\mu}{u}{v} = \vec{d} (G \setminus E (G_{\mu} ), u,v )$.
\end{definition}
The $F_2(\mu)$ intuitively represents the distance data when using the part of the $\kct$ shown in Figure~\ref{fig:SubgraphForF2}.
We can compute $F_2$ from the root of $\kct$ to the leaves. To describe how to compute $F_2 (\mu )$, let $\lambda$ be the parent node of $\mu$ in $\kct$.

If $\lambda = \rho$  (root node ), the edges of $G \setminus E (G_{\mu} )$ are only $\langle x_{\lambda}, y_{\lambda}\rangle,\langle y_{\lambda}, x_{\lambda}\rangle$, so each weight $\kdsVec{2}{\mu}{u}{v}$ can be calculated by replacing $\mu$ to $\lambda$ in $\kdsVec{1}{\mu}{u}{v}$ 
in Table~\ref{table:F1_Qnode}.

From here, we assume that $\lambda \neq \rho$. Then, $F_2 (\mu )$ can be calculated by using $H_{\lambda}\setminus \krefe{\mu} \cup \krefe{\lambda}$ instead of $G \setminus E (G_{\mu} )$. We set the weights of the edges $\langle x_\lambda,y_\lambda \rangle$ and $\langle y_\lambda,x_\lambda \rangle$ of this graph to $\kdsD{2}{\lambda}{x_\lambda}{y_\lambda}$ and $\kdsD{2}{\lambda}{y_\lambda}{x_\lambda}$, respectively.

For $\lambda \in S_{\kct}$, let $\kch{\lambda}=\{ \lambda_1, \ldots , \lambda_k \}, \mu =\lambda_i$, $x_{\lambda}=x_{\lambda_1}, y_{\lambda_{j}}=x_{\lambda_{j+1}}\  (1 \leq j \leq k-1 ), y_{\lambda_k}=y_{\lambda}$ in $\kskl{\lambda}$. Each weight $\kdsVec{2}{\mu}{u}{v}$ can be obtained in the same idea as in Table~\ref{table:F1_Snode}. We show here the formulas for calculating some of the weights.
\begin{align*}
    \kdsD{2}{\mu}{x_{\lambda_i}}{y_{\lambda_i}}
    &=\ksyx{\lambda}{1,i-1}+\kdsD{2}{\lambda}{x_\lambda}{y_\lambda}+\ksyx{\lambda}{i+1,k},\\
    \kdsBD{2}{\mu}{x_{\lambda_i}}{y_{\lambda_i}}
    &=\kdsD{2}{\mu}{x_{\lambda_i}}{y_{\lambda_i}}
    + \min \left\{ \begin{matrix}
        \displaystyle \min_{1\leq j \leq k, j \neq i} \left\{ \kbyx{\lambda}{j} \right\}\\ \kdsBD{2}{\lambda}{x_\lambda}{y_\lambda}-\kdsD{2}{\lambda}{x_\lambda}{y_\lambda}
    \end{matrix}\right\},
\end{align*}
\begin{align*}
    &\kdsBD{2}{\mu}{x_{\lambda_i}}{x_{\lambda_i}}\\
    &=\min \left\{\begin{matrix}
        \displaystyle \min_{1\leq j\leq i-1} \left\{ \kbyy{\lambda}{j}\right\} -(\ksyx{\lambda}{i,k}+\ksxy{\lambda}{i,k})\\
        \ksyx{\lambda}{1,i-1}+\kdsBD{2}{\lambda}{x_{\lambda}}{x_{\lambda}} +\ksxy{\lambda}{1,i-1}\\
        \displaystyle \ksyx{\lambda}{1,i-1}+\kdsD{2}{\lambda}{x_{\lambda}}{y_{\lambda}}+\min_{i+1\leq j\leq k} \left\{ \kbyy{\lambda}{j} \right\}+\kdsD{2}{\lambda}{y_{\lambda}}{x_{\lambda}} +\ksxy{\lambda}{1,i-1} 
    \end{matrix}\right\}.
\end{align*}

If $\lambda \in P_{\kct}$, it can be calculated in the same way as $F_1$ for P nodes, by noting that $x_\mu = x_\lambda, y_\mu = y_\lambda$. That is, each $\kdsVec{2}{\mu}{u}{v}$ can be calculated by the formula $\kdsVec{1}{\mu}{u}{v}$ in Table~\ref{table:F1_Pnode} if we set $\ell_{u,v}, \ell_{u,v}^{\mathrm{B}}$ as follows:
$$\ell_{u,v}=
\min \left\{ \begin{matrix}
    \kdsD{2}{\lambda}{u}{v}\\
    \min_{\theta \in \kch{\lambda} \setminus \{\mu\}} \left\{ \kdsD{1}{\theta}{u}{v} \right\}
\end{matrix}
 \right\}
,\ 
\ell^\mathrm{B}_{u,v}=
\min \left\{ \begin{matrix}
    \kdsBD{2}{\lambda}{u}{v}\\
    \min_{\theta \in \kch{\lambda} \setminus \{\mu\}} \left\{ \kdsBD{1}{\theta}{u}{v} \right\}
\end{matrix}  \right\}
.$$

If $\lambda \in R_{\kct}$, each weight $\kdsVec{2}{\mu}{u}{v}$ can be calculated on $H' \coloneqq H_{\lambda}\setminus \krefe{\mu} \cup \krefe{\lambda}$ as follows.
$$\kdsVec{2}{\mu}{u}{v}
=\begin{pmatrix}
    \kdis{H'}{u}{v}\\
    \min \left\{ \begin{matrix}
        \displaystyle \min_{p,q \in \{x_{\lambda},y_{\lambda}\}}\{ \kdis{H'}{u}{p} + \kdsBD{2}{\lambda}{p}{q} + \kdis{H'}{q}{v} \}\\
        \displaystyle \min_{\theta \in \kch{\lambda} \setminus \{\mu\}} \left\{ \min_{p,q \in \{x_{\theta},y_{\theta}\}}\{ \kdis{H'}{u}{p} + \kdsBD{1}{\theta}{p}{q} + \kdis{H'}{q}{v} \} \right\}
    \end{matrix} \right\}
\end{pmatrix}.
$$

Note that each $\kdis{H'}{a}{b}$ in the above equation is calculated by a shortest path algorithm for $H'$. An example of calculating $F_2$ is shown in Figure~\ref{fig:SPQRtree_F1F2} of Appendix.
\subsection{Definition of the mapping $F_3$ and its computation}
\begin{figure}[h]
    \centering
    \includegraphics[keepaspectratio, scale=0.5]{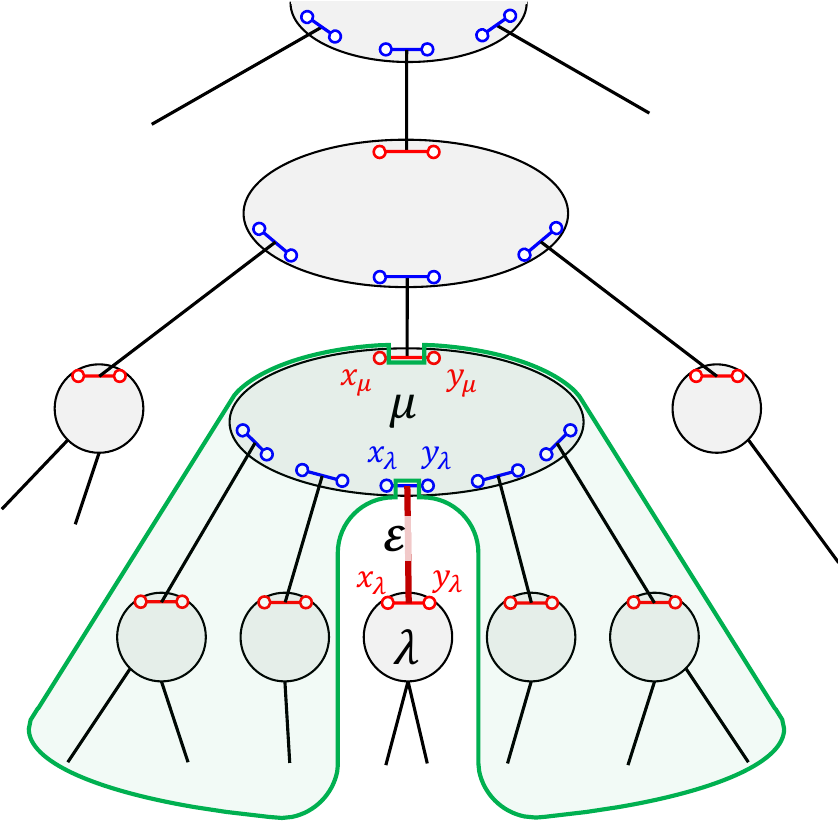}
    \caption{The subtree considered in  $F_3(\kce)$}
    \label{fig:SubgraphForF3}
\end{figure}

We define the mapping $F_3 \colon E (\kct )\setminus \{\{\rho, \rho'\}\} \to \kck$ as follows.
\begin{definition}\label{def:mapF3}
    For each edge $\kce= \{\mu,\lambda \} \in E (\kct )\setminus \{\{\rho, \rho'\}\}$  ($\lambda \in \kch{\mu}$ ), $F_3 (\kce )=K_{\mu,\lambda}^{\vec{f}_3 (\kce )}$ (a complete graph consists of at most 4 vertices). The weight of each vertex pair $\langle u,v \rangle \in  (\{x_{\mu}, y_{\mu}\} \cup \{x_{\lambda}, y_{\lambda}\} )^2$ is $\kdsVec{3}{\kce}{u}{v} = \vec{d} (G_{\mu}\setminus E (G_{\lambda} ), u,v )$.
\end{definition}
The $F_3(\kce)$ intuitively represents the distance data when using the part of the $\kct$ shown in Figure~\ref{fig:SubgraphForF3}.

In the actual $F_3 (\kce )$ calculation, we can consider $H_{\mu}\setminus \krefe{\lambda}$ instead of $G_{\mu}\setminus E (G_{\lambda} )$.

If $\mu \in S_{\kct}$, let $\kch{\mu}=\{ \mu_1, \ldots , \mu_k \}, \lambda=\mu_{i}$, and let $x_{\mu}=x_{\mu_1}, y_{\mu_{j}}=x_{\mu_{j+1}}\  (1 \leq j \leq k-1 ), y_{\mu_k}=y_{\mu}$ in $\kskl{\mu}$. Each weight $\kdsVec{3}{\kce}{u}{v}$ can be calculated as follows.

If $u,v \in \{x_{\mu}\}\cup \{x_{\mu_i}\}$, then $\kdsD{3}{\kce}{u}{v}=0$,
$$\kdsBD{3}{\kce}{u}{v}=\left\{ \begin{matrix} 0 & x_\mu \in B \\ \infty & x_\mu \notin B \end{matrix} \right.$$
if $i=1$  ($x_\mu = x_{\mu_i}$), and then $\kdsVec{3}{\kce}{u}{v}$ is shown in Table~\ref{table:F2_Snode} if $i\geq 2$  ($x_\mu \neq x_{\mu_i}$).

If $u,v \in \{y_{\mu}\}\cup \{y_{\mu_i}\}$, $\kdsVec{3}{\kce}{u}{v}$ can be calculated in the same way as above by considering the case when $i=k$  ($y_\mu = y_{\mu_i}$ ) and the case when $i\leq k-1$  ($y_\mu \neq y_{\mu_i}$) separately.

Otherwise, $u$ to $v$ cannot be reached in $H_\mu \setminus \krefe{\lambda}$, so $\kdsD{3}{\kce}{u}{v}=\kdsBD{3}{\kce}{u}{v}=\infty$.


If $\mu \in P_{\kct}$, it can be calculated in the same way as $F_1$ for P nodes, by noting that $x_\lambda = x_\mu, y_\lambda = y_\mu$. That is, each $\kdsVec{3}{\kce}{u}{v}$ can be calculated by the formula $\kdsVec{1}{\mu}{u}{v}$ in Table~\ref{table:F1_Pnode} if we set $\ell_{u,v}, \ell_{u,v}^{\mathrm{B}}$ as follows.
$$\ell_{u,v}=\min_{\theta \in \kch{\mu} \setminus \{\lambda\}} \left\{ \kdsD{1}{\theta}{u}{v} \right\},\ \ell^\mathrm{B}_{u,v}=\min_{\theta \in \kch{\mu} \setminus \{\lambda\}} \left\{ \kdsBD{1}{\theta}{u}{v} \right\}.$$

If $\mu \in R_{\kct}$, each weight $\kdsVec{3}{\kce}{u}{v}$ can be calculated on $H'' \coloneqq H_{\mu}\setminus \krefe{\lambda}$ as follows.
$$\kdsVec{3}{\kce}{u}{v}
=\begin{pmatrix}
    \kdis{H''}{u}{v}\\
    \min_{\theta \in \kch{\mu} \setminus \{\lambda \}} \{ \min_{p,q \in \{x_{\theta},y_{\theta}\}} \{ \kdis{H''}{u}{p} + \kdsBD{1}{\theta}{p}{q} + \kdis{H''}{q}{v} \} \}
\end{pmatrix}.
$$

\begin{figure}[h]
    \centering
    \includegraphics[keepaspectratio, scale=0.5]{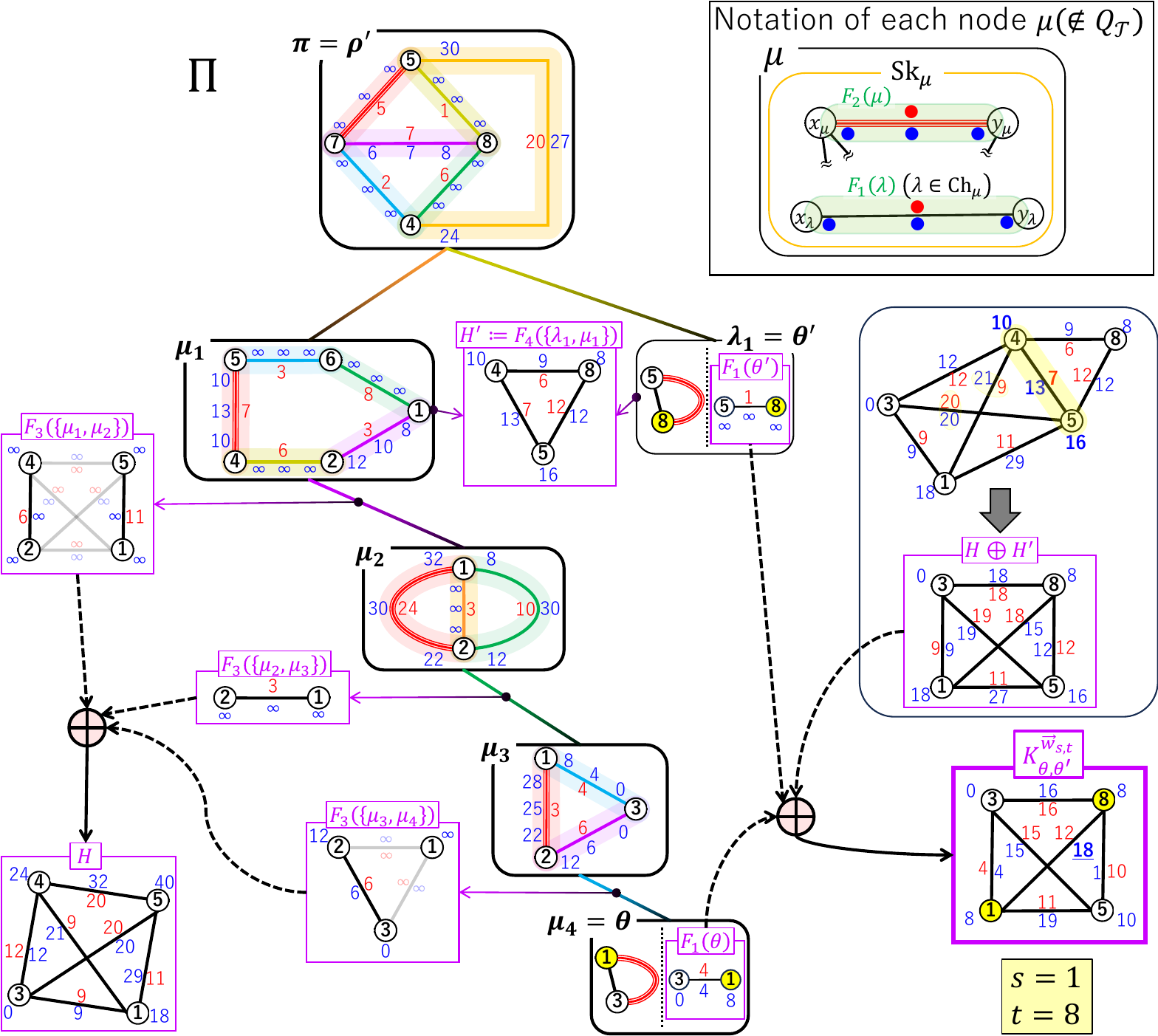}
    \caption{$F_3,F_4$ on $\Pi$ and calculation of the beer distance between $s=1$ and $t=8$}
    \label{fig:SPQRtree_F3F4}
\end{figure}

Note that each $\kdis{H''}{a}{b}$ in the above equation is calculated by the shortest path algorithm for $H''$. An example of calculating a part of $F_3$ is shown in Figure~\ref{fig:SPQRtree_F3F4}.

\subsection{Definition of the mapping $F_4$ and its computation}\label{subsec:defF4}
\begin{figure}[h]
    \centering
    \includegraphics[keepaspectratio, scale=0.5]{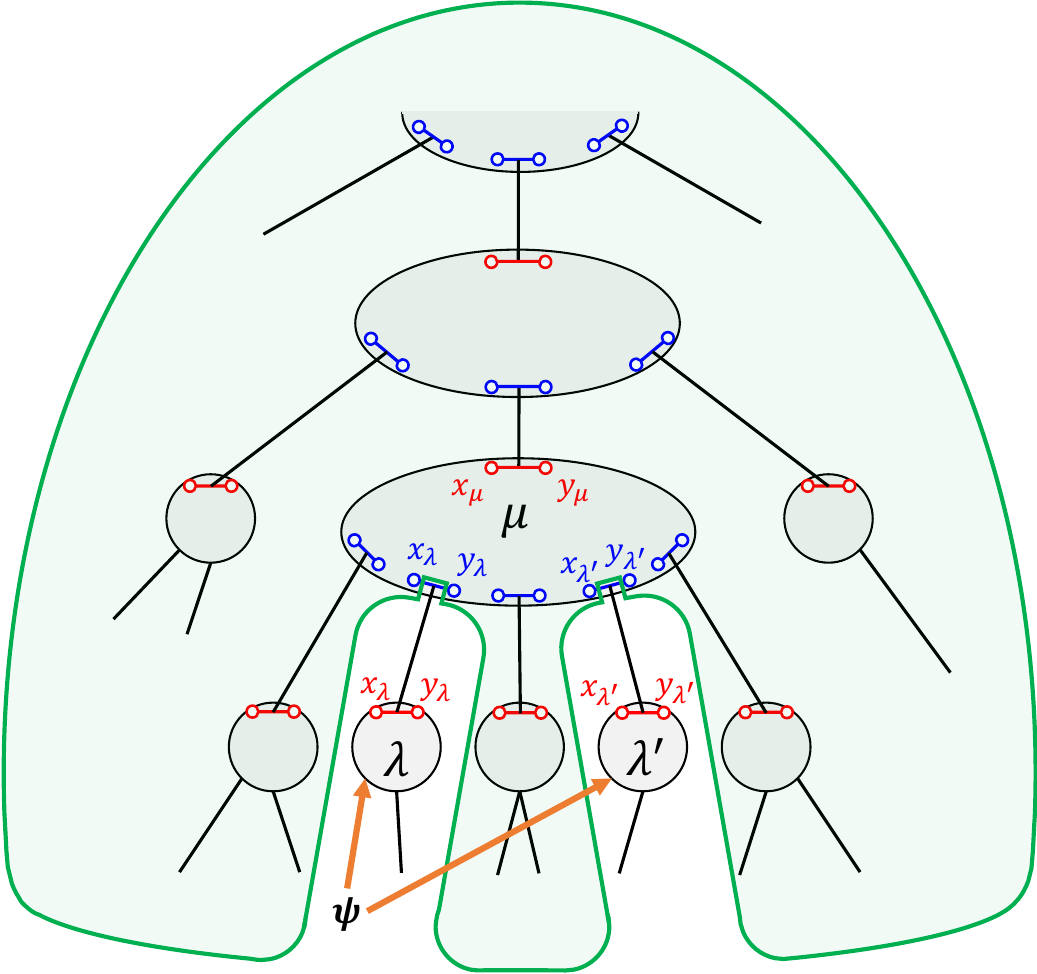}
    \caption{The subtree considered in  $F_4(\psi)$}
    \label{fig:SubgraphForF4}
\end{figure}

We define the mapping $F_4 \colon \bigcup_{\mu \in V (\kct )\setminus Q_\kct } 
\binom{\kch{\mu}}{2} \to \kck$ as follows.
\begin{definition}\label{def:mapF4}
    For each node $\mu \in V (\kct )\setminus Q_\kct $ and each node pair $ \psi = \{\lambda, \lambda'\} \in \binom{\kch{\mu}}{2}$ of $\mu$, $F_4 ( \psi )=K_{\lambda,\lambda'}^{\vec{f}_4 ( \psi )}$ (a complete graph consists of at most 4 vertices). The weight of each vertex pair $\langle u,v \rangle \in  (\{x_{\lambda}, y_{\lambda}\} \cup \{x_{\lambda'}, y_{\lambda'}\} )^2$ is $\kdsVec{4}{ \psi}{u}{v} = \vec{d} (G\setminus E (G_{\lambda} ) \setminus E (G_{\lambda'} ), u,v )$.
\end{definition}
The $F_4(\psi)$ intuitively represents the distance data when using the part of the $\kct$ shown in Figure~\ref{fig:SubgraphForF4}.
In the actual $F_4 ( \psi )$ calculation, we can consider $H_{\mu}\setminus \krefe{\lambda}\setminus \krefe{\lambda'} \cup \krefe{\mu}$ instead of $G\setminus E (G_{\lambda} ) \setminus E (G_{\lambda'} )$. We set the weights of the edges $\langle x_\mu,y_\mu \rangle, \langle y_\mu,x_\mu \rangle$ of this graph to $\kdsD{2}{\mu}{x_\mu}{y_\mu}, \kdsD{2}{\mu}{y_\mu}{x_\mu}$ respectively.

If $\mu \in S_{\kct}$, let $\kch{\mu}=\{ \mu_1, \ldots , \mu_k \}, \lambda=\mu_{i}, \lambda'=\mu_{j}$ ($i<j$), and let $x_{\mu}=x_{\mu_1}, y_{\mu_{p}}=x_{\mu_{p+1}}\  (1 \leq p \leq k-1 ), y_{\mu_k}=y_{\mu}$ in $\kskl{\mu}$. The weights $\kdsVec{4}{\psi}{u}{v}$ can be calculated in the same way as the weights of $F_3$ for the S nodes. We show here the formulas for calculating some of the weights.
\begin{align*}
    \kdsD{4}{\psi}{x_{\mu_i}}{y_{\mu_j}}
    &=\ksyx{\mu}{1,i-1}+\kdsD{2}{\mu}{x_\mu}{y_\mu}+\ksyx{\mu}{j+1,k},\\
    \kdsBD{4}{\psi}{x_{\mu_i}}{y_{\mu_j}}
    &=\kdsD{4}{\psi}{x_{\mu_i}}{y_{\mu_j}} +\min \left\{ \begin{matrix}
        \kdsBD{3}{\{\mu,\lambda\}}{x_{\mu_i}}{x_\mu}-\ksyx{\mu}{1,i-1}\\
        \kdsBD{2}{\mu}{x_\mu}{y_\mu}-\kdsD{2}{\mu}{x_\mu}{y_\mu}\\
        \kdsBD{3}{\{\mu,\lambda'\}}{y_\mu}{y_{\mu_j}}-\ksyx{\mu}{j+1,k}
    \end{matrix} \right\},
\end{align*}
\begin{align*}
    &\kdsBD{4}{\psi}{x_{\mu_i}}{x_{\mu_i}}\\
    &=\min \left\{\begin{matrix}
        \kdsBD{3}{\{\mu,\lambda\}}{x_{\mu_i}}{x_{\mu_i}}\\
        \ksyx{\mu}{1,i-1} +\kdsBD{2}{\mu}{x_\mu}{x_\mu}+\ksxy{\mu}{1,i-1}\\
        \ksyx{\mu}{1,i-1} +\kdsD{2}{\mu}{x_\mu}{y_\mu}+\kdsBD{3}{\{\mu,\lambda'\}}{y_\mu}{y_\mu}+ \kdsD{2}{\mu}{y_\mu}{x_\mu}+\ksxy{\mu}{1,i-1}
    \end{matrix}\right\},
\end{align*}
\begin{align*}
    \kdsD{4}{\psi}{y_{\mu_i}}{x_{\mu_j}}
    &=\ksxy{\mu}{i+1,j-1},\\
    \kdsBD{4}{\psi}{y_{\mu_i}}{x_{\mu_j}}
    &=\left\{\begin{matrix}
        0 & j=1+1, y_{\mu_i}\in B\\
        \infty & j=1+1, y_{\mu_i}\notin B\\
        \displaystyle \kdsD{4}{\psi}{y_{\mu_i}}{x_{\mu_j}}+ \min_{i+1\leq p\leq j-1} \left\{ \kbxy{\mu}{p}\right\} & j\geq i+2
    \end{matrix}\right. ,\\
    \kdsBD{4}{\psi}{y_{\mu_i}}{y_{\mu_i}}
    &=\left\{\begin{matrix}
        0 & j=1+1, y_{\mu_i}\in B\\
        \infty & j=1+1, y_{\mu_i}\notin B\\
        \displaystyle \min_{i+1\leq p\leq j-1} \left\{ \kbxx{\mu}{p}\right\} -(\ksxy{\mu}{1,i}+\ksyx{\mu}{1,i}) & j\geq i+2
    \end{matrix}\right. .
\end{align*}

If $\mu \in P_{\kct}$, it can be calculated in the same way as $F_1$ for P nodes, by noting that $x_{\lambda} = x_{\lambda'} = x_\mu, y_{\lambda} = y_{\lambda'} = y_\mu$. That is, each $\kdsVec{4}{ \psi}{u}{v}$ can be calculated by the formula $\kdsVec{1}{\mu}{u}{v}$ in Table~\ref{table:F1_Pnode} if we set $\ell_{u,v}, \ell_{u,v}^{\mathrm{B}}$ as follows.
$$\ell_{u,v}=
\min \left\{ \begin{matrix}
    \kdsD{2}{\mu}{u}{v} \\
    \displaystyle \min_{\theta \in \kch{\mu} \setminus \{\lambda, \lambda' \}} \left\{ \kdsD{1}{\theta}{u}{v} \right\} 
\end{matrix} \right\}
,\ \ell^\mathrm{B}_{u,v}=
\min \left\{ \begin{matrix}
    \kdsD{2}{\mu}{u}{v}\\
    \displaystyle \min_{\theta \in \kch{\mu} \setminus \{\lambda, \lambda' \}} \left\{ \kdsBD{1}{\theta}{u}{v} \right\}
\end{matrix} \right\}
.$$
Note that if $\kch{\mu}=\{\lambda, \lambda'\}$, then $\ell_{u,v}= \kdsD{2}{\mu}{u}{v}, \ell^\mathrm{B}_{u,v}=\kdsD{2}{\mu}{u}{v}$.

If $\mu \in R_{\kct}$, each weight $\kdsVec{4}{ \psi}{u}{v}$ can be calculated on $H''' \coloneqq H_{\mu}\setminus \krefe{\lambda} \setminus \krefe{\lambda'} \cup \krefe{\mu}$ as follows.
$$\kdsVec{4}{\psi}{u}{v}
=\begin{pmatrix}
    \kdis{H'''}{u}{v}\\
    \min \left\{ \begin{matrix}
        \displaystyle \min_{p,q \in \{x_{\mu},y_{\mu}\}} \{ \kdis{H'''}{u}{p} + \kdsBD{2}{\mu}{p}{q} + \kdis{H'''}{q}{v} \}\\
        \displaystyle \min_{\theta \in \kch{\mu} \setminus \{\lambda, \lambda' \}} \left\{ \min_{p,q \in \{x_{\theta},y_{\theta}\}}\{ \kdis{H'''}{u}{p} + \kdsBD{1}{\theta}{p}{q} + \kdis{H'''}{q}{v} \} \right\}
    \end{matrix} \right\}
\end{pmatrix}.
$$

Note that each $\kdis{H'''}{a}{b}$ in the above equation is calculated by the shortest path algorithm for $H'''$. An example of calculating an image of $F_4$ is shown in Figure~\ref{fig:SPQRtree_F3F4}.

Here, if $\mu\in S_\kct \cup P_\kct$, $F_4 ( \psi )$ can be computed in $O(1)$ time by using Range Minimum Query. Also, the beer distance is obtained from a graph that is a combination of the images of $F_1,F_2,F_3,F_4$, but the image of $F_4$ appears in at most one element of the combination (see subsection \ref{subsec:BeerDistanceRepresentation} for details). Therefore, it is enough to compute $F_4$ only for the child node pairs of the R node in the preprocessing. From this it is convenient to consider a mapping restricting the domain of $F_4$ and we define $F_{4\mathrm{R}} \colon \bigcup_{\mu \in R_\kct } 
\binom{\kch{\mu}}{2} \to \kck$  ($F_{4\mathrm{R}} ( \psi ) = F_4 ( \psi )$, $ \psi \in \bigcup_{\mu \in R_\kct } 
\binom{\kch{\mu}}{2}$).
Because of space limitation, we show analyses on computational complexities in Section~\ref{sec:complexity}.

\section{Algorithm based on triconnected component decomposition}\label{sec:trialgo}
\subsection{Definition of binary operations}
We define the binary operation $\oplus \colon \kck^2 \to \kck $ as follows.

\begin{definition}
For each $H_1,H_2 \in \kck$, $H_1 \oplus H_2$ is defined as follows. If $H_1 = \perp$ or $H_2 = \perp$, $H_1 \oplus H_2 =\perp$. If $H_1 \neq \perp$ and $H_2 \neq \perp$, let $H_i = K_{\mu_i,\lambda_i}^{\vec{w}_i}$ ($i=1,2$). Also, let $A=(\{\mu_1\}\cup\{\lambda_1\})\cap (\{\mu_2\}\cup\{\lambda_2\})$ be the set of nodes in $\kck$ that give vertices appearing in $H_1,H_2$ in common.

If $|A|\neq 1$, $H_1 \oplus H_2 =\perp$. If $|A|=1$, let $A=\{\theta\}$, $H_1 \oplus H_2 = K_{\theta_1,\theta_2}^{\vec{w}}$. Here, we define $\theta_1,\theta_2$ as follows.
$$\theta_1=\left\{\begin{matrix}
    \mu_1 (=\lambda_1 ) & \mu_1 =\theta = \lambda_1\\
    \mu_1 & \mu_1\neq \theta =\lambda_1\\
    \lambda_1 & \mu_1 = \theta \neq \lambda_1
    \end{matrix}\right.,\ \
\theta_2=\left\{\begin{matrix}
    \mu_2 (=\lambda_2 ) & \mu_2 =\theta = \lambda_2\\
    \mu_2 & \mu_2\neq \theta =\lambda_2\\
    \lambda_2 & \mu_2 = \theta \neq \lambda_2
    \end{matrix}\right..$$
Also, let $H_1 \Tilde{\cup} H_2$ be a weighted multi graph with vertex set $V(H_1)\cup V(H_2)$, given distinct edges in $H_1$ and edges in $H_2$, and let $\vec{z}$ be the weights defined by
$$\vec{z}(e)=\begin{pmatrix}z(e)\\ z^\mathrm{B}(e)\end{pmatrix}=\vec{w}_i(p,q) \ (e=\langle p,q \rangle \in E(H_i), i=1,2).$$
Then, For each $\langle u,v \rangle \in  (\{x_{\theta_1},y_{\theta_1}\}\cup \{x_{\theta_2},y_{\theta_2}\} )^2$, we define $\vec{w}(u,v) $ as follows.
$$w(u,v) = \kdis{(H_1 \Tilde{\cup} H_2 ,z)}{u}{v},$$
$$w^\mathrm{B}(u,v) = \min_{e=\langle p,q \rangle \in E(H_1 \Tilde{\cup} H_2)} \{\kdis{(H_1 \Tilde{\cup} H_2 ,z)}{u}{p} + z^\mathrm{B}(e)+\kdis{(H_1 \Tilde{\cup} H_2 ,z)}{q}{v}\}.$$
\end{definition}
For a concrete example of this operation, see the computation of $H \oplus H'$ in Figure~\ref{fig:SPQRtree_F3F4}. Furthermore, we define a subset $\widehat{\kck}$ of $\kck$ by
$$\widehat{\kck} = \{ K_{\mu,\lambda}^{\vec{w}} \in \kck \setminus \{\perp\} \mid \lambda \in \kdes{\mu} \} \cup \{\perp\}$$
and define the binary operation $\widehat{\oplus} \colon \widehat{\kck}^2 \to \widehat{\kck} $ as follows.
\begin{definition}
    For each $H_1,H_2 \in \widehat{\kck}$, if $H_1 = \perp$ or $H_2 = \perp$, then $H_1 \widehat{\oplus} H_2 =\perp$, otherwise, let $H_i = K_{\mu_i,\lambda_i}^{\vec{w}_i}, \lambda_i \in \kdes{\mu_i}$ ($i=1,2$) then
    $$H_1 \widehat{\oplus} H_2 =\left\{\begin{matrix}
        \perp & \lambda_1 \neq \mu_2 \\
        H_1 \oplus H_2 & \lambda_1 = \mu_2
    \end{matrix}\right. .$$
\end{definition}

\begin{lemma}\label{lemma:semigroup}
    $\widehat{\oplus}$ is a semigroup.
\end{lemma}
The proof is given in Section~\ref{sec:proofofsemigroup}.

\subsection{Representation of distance and beer distance using mapping and algorithms for Beer Path Query}\label{subsec:BeerDistanceRepresentation}

By using $F_1, F_2, F_3, F_4, F_{4\mathrm{R}}$ and tree product query data structures, we
can compute beer path between given vertices. Recall that $r$ is the maximum edge number of the skeleton of R nodes in the SPQR tree of $G$ and $W$ is the range of the edge weight function.

\begin{theorem}
    If we precompute $F_1,F_2$ as a data structure, the space required to store the data structure is $O(m)$. The preprocessing time and query time are
    \begin{enumerate}
        \item $O(m+r\cdot\min\{m,rn\})$ and $O(n+r\cdot\min\{m,rn\}+\alpha(m))$ if $W=\mathbb{Z}_{\geq 0}$ and $G$ is undirected.
        \item $O(m+r(m+n\log r_+))$ and $O(n+ r(m+n\log r_+) +\alpha(m))$ if $W=\mathbb{R}_{\geq 0}$.
    \end{enumerate}
\end{theorem}

\begin{theorem}
    If we precompute $F_1,F_2,F_3$ as a data structure, the space required to store the data structure is $O(m)$. The preprocessing time and query time are
    \begin{enumerate}
        \item $O(m+r^2\cdot\min\{m,rn\})$ and $O(r^2+\alpha(m))$ if $W=\mathbb{Z}_{\geq 0}$ and $G$ is undirected.
        \item $O(m+r^2(m+n\log r_+))$ and $O(r^2\log r_+ +\alpha(m))$ if $W=\mathbb{R}_{\geq 0}$.
    \end{enumerate}
\end{theorem}

\begin{theorem}
    If we precompute $F_1,F_2,F_3,F_{4\mathrm{R}}$ as a data structure, the space required to store the data structure is $O(m+r\cdot\min\{m,rn\})$. The preprocessing time and query time are
    \begin{enumerate}
        \item $O(m+r^3\cdot\min\{m,rn\})$ and $O(\alpha(m))$ if $W=\mathbb{Z}_{\geq 0}$ and $G$ is undirected.
        \item $O(m+r^3(m+n\log r_+))$ and $O(\alpha(m))$ if $W=\mathbb{R}_{\geq 0}$.
    \end{enumerate}
\end{theorem}
Proofs are given in Section~\ref{sec:proofsofalgorithms}

\bibliography{bp}

\newpage

\appendix

\section{Missing Proofs}

\subsection{Proof of Lemma~\ref{lemma:semigroup}}\label{sec:proofofsemigroup}
\begin{figure}[h]
    \centering
    \includegraphics[keepaspectratio, scale=0.35]{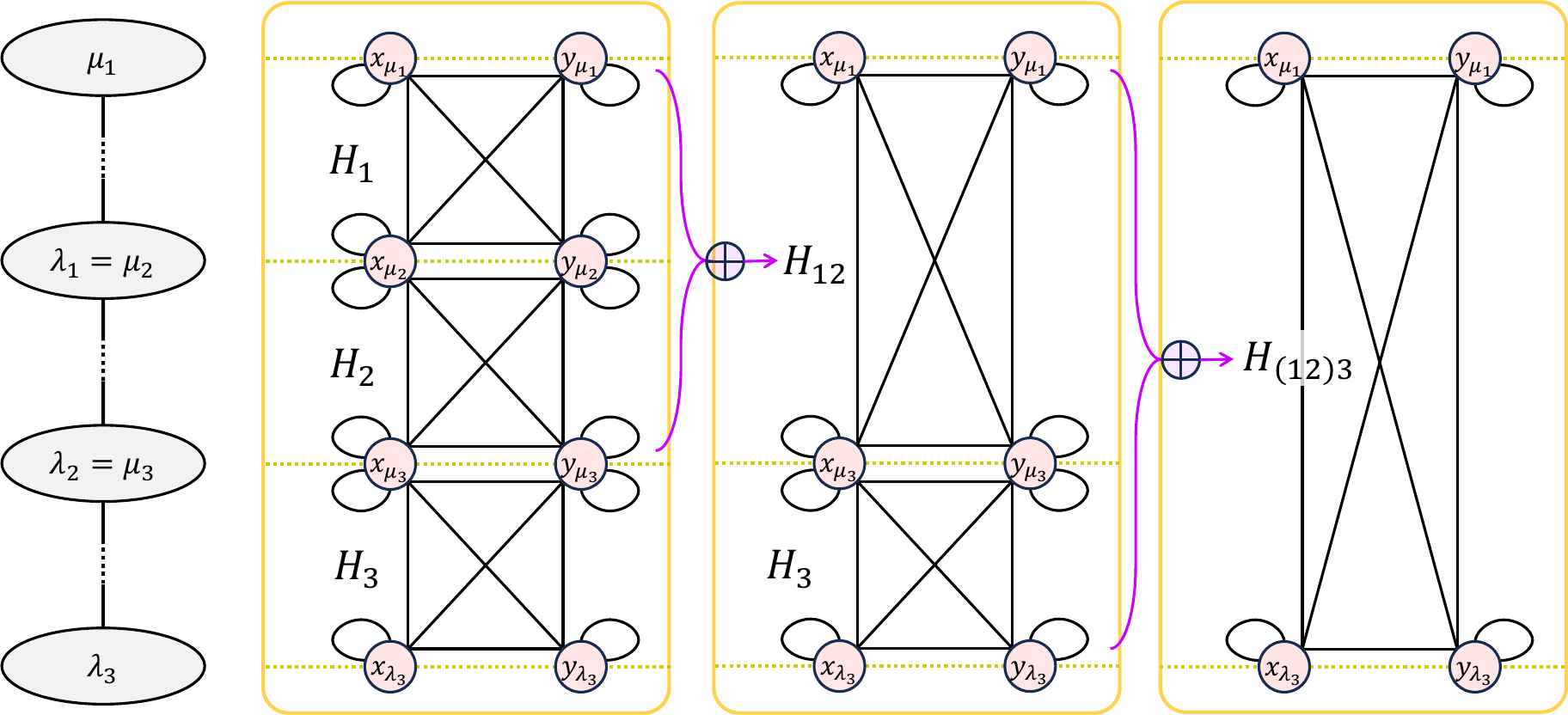}
    \caption{Calculation of $H_{(12)3}$ in the case of $\lambda_1=\mu_2$ and $\lambda_2=\mu_3$}
    \label{fig:semigroup}
\end{figure}

\begin{proof}
    We arbitrarily take $H_1,H_2,H_3\in \widehat{\kck}$, and confirm that $H_{(12)3} \coloneqq (H_1 \widehat{\oplus} H_2)\widehat{\oplus} H_3$ and $H_{1(23)}\coloneqq H_1 \widehat{\oplus}(H_2 \widehat{\oplus} H_3)$ are equal. First, if $H_1= \perp$ or $H_2= \perp$ or $H_3= \perp$, clearly $H_{(12)3}=H_{1(23)}=\perp$. In the following, let $H_i\neq \perp$ and $H_i=K_{\mu_i,\lambda_i}^{\vec{w}_i}$ ($\lambda_i \in \kdes{\mu_i}$) for each $i$. If $\lambda_1 \neq \mu_2$ or $\lambda_2 \neq \mu_3$, then we can easily obtain $H_{(12)3}=H_{1(23)}=\perp$. If $\lambda_1 = \mu_2,\lambda_2 =\mu_3$, let $H_{(12)3}=K_{\mu_1,\lambda_3}^{\vec{w}_{(12)3}},H_{1(23)}=K_{\mu_1,\lambda_3}^{\vec{w}_{1(23)}}$. Then, we show briefly that $\vec{w}_{(12)3}(u,v)=\vec{w}_{1(23)}(u,v)$ for each $u,v\in \{x_{\mu_1},y_{\mu_1}\}\cup \{x_{\lambda_3},y_{\lambda_3}\}$.

    If let $H_1 \widehat{\oplus}H_2=H_{12}=K_{\mu_1,\mu_3}^{\vec{w}_{12}}$, then from Figure~\ref{fig:semigroup} the following holds for each $u'\in \{x_{\mu_1},y_{\mu_1}\}$ and $v' \in \{x_{\mu_3},y_{\mu_3}\}$.
    $$w_{12}(u',v')=\kdis{H_{12}}{u'}{v'}=\min_{p\in \{x_{\mu_2},y_{\mu_2}\}} \{\kdis{H_1}{u'}{p} + \kdis{H_2}{p}{v'}\}.$$
    By using this, the following holds for each $u\in \{x_{\mu_1},y_{\mu_1}\}$ and $v \in \{x_{\lambda_3},y_{\lambda_3}\}$.
    \begin{align*}
        w_{(12)3}(u,v)
        &=\kdis{H_{(12)3}}{u}{v}=\min_{q\in\{x_{\mu_3},y_{\mu_3}\}} \{\kdis{H_{12}}{u}{q}+\kdis{H_3}{q}{v}\}\\
        &=\min_{q\in\{x_{\mu_3},y_{\mu_3}\}} \left\{\min_{p\in \{x_{\mu_2},y_{\mu_2}\}} \{\kdis{H_1}{u}{p} + \kdis{H_2}{p}{q}\}+\kdis{H_3}{q}{v}\right\}\\
        &=\min_{\substack{p\in \{x_{\mu_2},y_{\mu_2}\}\\q\in\{x_{\mu_3},y_{\mu_3}\}}} \{\kdis{H_1}{u}{p} + \kdis{H_2}{p}{q}+\kdis{H_3}{q}{v}\}.
    \end{align*}
    By the same idea, exactly the same result is obtained for $w_{1(23)}(u,v)$. We can show $w_{(12)3}(u,v)=w_{1(23)}(u,v)$ and $w_{(12)3}^{\mathrm{B}}(u,v)=w_{1(23)}^{\mathrm{B}}(u,v)$ for the other weights in the same way.
\end{proof}

\subsection{Preprocessing algorithms}

\subsubsection{Algorithm for preprocessing $F_1,F_2$}
We consider an algorithm that preprocesses $F_1,F_2$. For this algorithm, the preprocessing space is $C_1^{\mathrm{space}}+C_2^{\mathrm{space}}=O (m)$ and the preprocessing time is
\[
C_1^{\mathrm{time}}+C_2^{\mathrm{time}}=\left\{
\begin{matrix}
    O (m+r\cdot \min \{m,rn\} ) & W=\mathbb{Z}_{\geq 0}\\
    O (m+r (m+n \log r_{+} ) ) & W=\mathbb{R}_{\geq 0}.
\end{matrix}
\right.
\]
Beer Path Query can be solved by computing $O(n)$ images of $F_3$ and one image of $F_4$ on $\Pi$ and combine them using Tree Product Query.
Thus, query time is 
\begin{align*}
    &O \left(\sum_{\mu \in S_\kct\cup P_\kct}1+\sum_{\mu \in R_\kct}m_\mu \kAlgD{H_\mu}+m_\pi \kAlgD{H_\pi} + \alpha (m) \right)\\ 
    &=\left\{\begin{matrix}
    O (n+r\cdot \min \{m,rn\}+ \alpha(m)) & W=\mathbb{Z}_{\geq 0} \\
    O (n+r (m+n \log r_{+} )+ \alpha(m)) & W=\mathbb{R}_{\geq 0}. 
\end{matrix}
\right.
\end{align*}

\subsubsection{Algorithm for preprocessing $F_1,F_2,F_3$}
We consider an algorithm that preprocesses $F_1,F_2,F_3$. For this algorithm, the preprocessing space is $C_1^{\mathrm{space}}+C_2^{\mathrm{space}}+C_3^{\mathrm{space}}=O (m)$ and the preprocessing time is
\[C_1^{\mathrm{time}}+C_2^{\mathrm{time}}+C_3^{\mathrm{time}}=\left\{
\begin{matrix}
    O (m+r^2\cdot \min \{m,rn\} ) & W=\mathbb{Z}_{\geq 0} \\
    O (m+r^2 (m+n \log r_{+} ) ) & W=\mathbb{R}_{\geq 0}.
\end{matrix}
\right.
\]
Beer Path Query can be solved by computing one image of $F_4$ and combine images of $F_3$ on $\Pi$ and $F_4$ using Tree Product Query.
Thus, query time is 
$$O (m_\pi \kAlgD{H_\pi} + \alpha (m ) )=\left\{\begin{matrix}
    O (r^2 +\alpha (m ) )
 & W=\mathbb{Z}_{\geq 0} \\
    O (r^2 \log r_{+} +\alpha (m ) ) & W=\mathbb{R}_{\geq 0}.
\end{matrix}\right.$$

\subsubsection{Algorithm for preprocessing $F_1,F_2,F_3,F_{4\mathrm{R}}$}
We consider an algorithm that preprocesses $F_1,F_2,F_3,F_{4\mathrm{R}}$. For this algorithm, the preprocessing space is $C_1^{\mathrm{space}}+C_2^{\mathrm{space}}+C_3^{\mathrm{space}}+C_{4\mathrm{R}}^{\mathrm{space}}=O (m+r\cdot \min \{m,rn\} )$ and the preprocessing time is 
\[C_1^{\mathrm{time}}+C_2^{\mathrm{time}}+C_3^{\mathrm{time}}+C_{4\mathrm{R}}^{\mathrm{time}}=\left\{
\begin{matrix}
    O (m+r^3\cdot \min \{m,rn\} ) & W=\mathbb{Z}_{\geq 0} \\
    O (m+r^3 (m+n \log r_{+} ) ) & W=\mathbb{R}_{\geq 0}.
\end{matrix}
\right.
\]
Beer Path Query can be solved by combining images of $F_3$ and  $F_4$ on $\Pi$ using Tree Product Query.
Thus, query time is $O (\alpha (m ) )$.

\subsection{Computational complexity for each mapping}\label{sec:complexity}
In this subsection, we analyze the computational complexity for each mapping.
Let $C_i^\mathrm{time}$ and $C_i^\mathrm{space}$ be the time required to compute each $F_i$ and the space to store, respectively.
\subsubsection{Computational complexity for $F_1$}
First, we consider $C_1^\mathrm{space}$. For each $\mu \in V (\kct )\setminus \{\rho\}$, $F_1 (\mu )$ is a graph of constant size. Also, each $\mu \in S_\kct \cup P_\kct $ uses $O(m_\mu)$ space in the preprocessing for Range Minimum Query and so on. Thus, $C_1^\mathrm{space}=O\left( \sum_{\mu \in Q_\kct \cup R_\kct} 1+ \sum_{\mu \in S_\kct \cup P_\kct} m_\mu \right)=O(m)$.

Next, we consider $C_1^\mathrm{time}$. If $\mu \in Q_\kct \setminus \{\rho \}$, $F_1(\mu)$ can be computed in $O(1)$ time. If $\mu \in S_\kct \cup P_\kct$, preprocessing and $F_1(\mu)$ calculation can be done in $O(m_\mu)$ time. Also, if $\mu \in R_\kct$, $F_1 (\mu )$ can be obtained by running the shortest path algorithm for $H_\mu$ for $O (m_\mu )$ times, so it can be computed in $O\left (m_{\mu} \kAlgD{H_\mu}\right )$ time. Thus, noting the definition of $r$, $C_1^\mathrm{time}$ is as follows: 
\begin{align*}
    C_1^\mathrm{time}
&= O\left ( \sum_{\mu \in Q_\kct} 1 + \sum_{\mu \in S_\kct \cup P_\kct} m_\mu + \sum_{\mu \in R_\kct} m_\mu \kAlgD{H_\mu} \right ) \\
&=O\left ( m + m + r \sum_{\mu \in R_\kct} \kAlgD{H_\mu}\right )
= \left\{
\begin{matrix}
    O (m+r\cdot \min \{m,rn\} ) & W=\mathbb{Z}_{\geq 0} \\
    O (m+r (m+n \log r_{+} ) ) & W=\mathbb{R}_{\geq 0}
\end{matrix}
\right. .
\end{align*}
\subsubsection{Computational complexity for $F_2$}
First, $C_2^\mathrm{space}$ can be similarly considered to $C_1^\mathrm{space}$, $C_2^\mathrm{space}=\sum_{\mu \in V (\kct )\setminus \{\rho\}} O (1)=O(m)$. Next, we consider $C_2^{\mathrm{time}}$. Let $\lambda$ be the parent node of $\mu$ in $\kct$. If $\lambda \in \{\rho\} \cup S_\kct \cup P_\kct$, $F_2 (\mu )$ can be computed in $O (1)$ time by using Table~\ref{table:F1_Qnode} or Range Minimum Query. Also, if $\lambda \in R_\kct$, $F_2 (\mu )$ can be obtained by running the shortest path algorithm for $H_{\lambda} \setminus \krefe{\mu} \cup \krefe{\lambda}$ for $O(m_\lambda)$ times, so it can be computed in $O\left (m_{\lambda} \kAlgD{H_\lambda}\right )$ time.
Thus, $C_2^\mathrm{time}$ can be evaluated as follows:
\begin{align*}
    C_2^\mathrm{time}
&= O\left ( \sum_{\lambda \in \{\rho\} \cup S_\kct \cup P_\kct} 1 + \sum_{\lambda \in R_\kct} m_\lambda \kAlgD{H_\lambda} \right )
=O\left (n + r \sum_{\lambda \in R_\kct} \kAlgD{H_\lambda}\right )\\
&= \left\{
\begin{matrix}
    O (n+r\cdot \min \{m,rn\} ) & W=\mathbb{Z}_{\geq 0}, \\
    O (n+r (m+n \log r_{+} ) ) & W=\mathbb{R}_{\geq 0}.
\end{matrix}
\right.
\end{align*}
\subsubsection{Computational complexity for $F_3$}
First, we consider $C_3^\mathrm{space}$. In $F_3$, a graph of a constant size is prepared for each edge of $E (\kct )\setminus \{\{\rho, \rho'\}\} $, so $C_3^\mathrm{space}=O (|E (\kct )| )=O (m )$.

Next, we consider $C_3^{\mathrm{time}}$. For each $\kce =\{\mu,\lambda\} \in E (\kct)\setminus \{\{\rho, \rho'\}\}$ ($\lambda\in \kch{\mu}$), if $\mu \in S_\kct \cup P_\kct$ then $F_3 (\kce)$ can be computed in $O(1)$ time by using Range Minimum Query. If $\mu \in R_\kct$ then $F_3 (\kce )$ can be obtained by running the shortest path algorithm for $H_\mu \setminus \krefe{\lambda}$ for $O (m_\lambda )$ times, so it can be computed in $O\left (m_{\lambda} \kAlgD{H_\lambda}\right )$ time.
Thus, $C_3^\mathrm{time}$ can be evaluated as follows.
\begin{align*}
    C_3^\mathrm{time}
&= O\left ( \sum_{\mu \in S_\kct \cup P_\kct} \sum_{\lambda\in \kch{\mu}}1 + \sum_{\mu \in R_\kct} \sum_{\lambda\in \kch{\mu}} m_\mu \kAlgD{H_\mu} \right ) \\
&=O\left ( m + r^2 \sum_{\mu \in R_\kct} \kAlgD{H_\mu}\right )
= \left\{
\begin{matrix}
    O (m+r^2 \cdot \min \{m,rn\} ) & W=\mathbb{Z}_{\geq 0}, \\
    O (m+r^2  (m+n \log r_{+} ) ) & W=\mathbb{R}_{\geq 0}.
\end{matrix}
\right.
\end{align*}
\subsubsection{Computational complexity for $F_4$}
If $F_4$ is realized as a data structure, it is not efficient because it requires computation and space even for images that can be computed in $O (1 )$ time, as described in Subsection \ref{subsec:defF4}. Therefore, we consider the computational complexity of realizing $F_{4\mathrm{R}}$ as a data structure instead of $F_4$ itself.

First, the space for $F_{4\mathrm{R}}$ is $C_{4\mathrm{R}}^\mathrm{space}=\sum_{\mu \in R_\kct}O\left ( m_{\mu}^2\right )=O (r\cdot \min \{m,rn\} )$. Next, we consider the preprocessing time of $F_{4\mathrm{R}}$, $C_{4\mathrm{R}}^{\mathrm{space}}$. For each $\mu \in R_\kct$ and each node pair $ \psi =\{\lambda,\lambda'\}\in \binom{\kch{\mu}}{2}$, $F_{4\mathrm{R}} ( \psi )=F_4 ( \psi )$ can be obtained by running the shortest path algorithm for $H_\mu \setminus \krefe{\lambda} \setminus \krefe{\lambda'} \cup \krefe{\mu}$ for $O (m_\mu )$ times, so it can be computed in $O\left (m_{\mu} \kAlgD{H_\mu}\right )$ time. Thus, $C_{4\mathrm{R}}^\mathrm{time}$ is as follows:
\begin{align*}
    C_{4\mathrm{R}}^\mathrm{time}
&= O\left ( \sum_{\mu \in R_\kct} \sum_{  \psi \in \binom{\kch{\mu}}{2}} m_\mu \kAlgD{H_\mu} \right )
= O\left ( \sum_{\mu \in R_\kct} m_{\mu}^3 \kAlgD{H_\mu} \right )\\
&=O\left ( r^3 \sum_{\mu \in R_\kct} \kAlgD{H_\mu}\right )
= \left\{
\begin{matrix}
    O (r^3 \cdot \min \{m,rn\} ) & W=\mathbb{Z}_{\geq 0}, \\
    O (r^3  (m+n \log r_{+} ) ) & W=\mathbb{R}_{\geq 0}.
\end{matrix}
\right.
\end{align*}

\subsection{Proofs for the algorithms}\label{sec:proofsofalgorithms}
In the following, we describe an outline of the algorithm corresponding to each of the above theorems.

First, we describe the representation of distances and beer distances using each mapping $F_i$ and binary operations. We also show several algorithms for Beer Path Query based on them. We consider computing the distance or beer distance from $s$ to $t$ in a biconnected connected graph $G$. First, let $\theta \in Q_{\kct}\setminus \{\rho\}$ be a Q node whose skeleton contains vertex $s$, and let $\{x_{\theta},y_{\theta}\}=\{s,s'\}$. Similarly, take a node $\theta' \in Q_{\kct}\setminus \{\rho\}$ whose skeleton contains a vertex $t$ and let $\{x_{\theta'},y_{\theta'}\}=\{t,t'\}$.

If $\theta = \theta'$, then we combine $F_1 (\theta)$ which contains data on the distance and the beer distance in $G_{\theta}$, and $F_2 (\theta)$ which contains data in $G\setminus E (G_{\theta})$. The combined result is represented by $F_1(\theta)\oplus F_2(\theta)$, and the distance and beer distance are obtained by referring to the weights of the vertex pair $\langle s,t \rangle$ in $F_1(\theta)\oplus F_2(\theta)$.

If $\theta \neq \theta'$, let $\pi$ be the lowest common ancestor of $\theta$ and $\theta'$ in $\kct$ and denote the $\theta$-$\theta'$ path in $\kct$ by the following vertex sequence $\Pi$.
$$\Pi \colon \theta = \mu_k , \mu_{k-1}, \ldots , \mu_2, \mu_1 ,\pi , \lambda_1 ,\lambda_2 ,\ldots , \lambda_{\ell-1}, \lambda_\ell = \theta'.$$
$F_1 (\theta)=F_1 (\mu_k )$ contains the data of the distance and the beer distance of each pair of $\{s,s'\}=\{x_{\mu_{k}}, y_{\mu_{k}}\}$ in $G_{\mu_k}$. Also, $F_3 (\{\mu_{k-1},\mu_{k}\} )$ contains the data of each pair of $\{x_{\mu_{k-1}}, y_{\mu_{k-1}}\}\cup\{x_{\mu_{k}}, y_{\mu_{k}}\}$ in $G_{\mu_{k-1}}\setminus E (G_{\mu_{k}} )$. Therefore, by combining $F_1 (\mu_k )$ and $F_3 (\{\mu_{k-1},\mu_{k}\} )$, we can obtain the data of each pair of $\{x_{\mu_{k-1}}, y_{\mu_{k-1}}\}\cup\{s,s'\}$ in $G_{\mu_{k-1}}$. And the combined result can be expressed as $F_1 (\mu_k )\oplus F_3 (\{\mu_{k-1},\mu_{k}\} )$.

By applying this idea repeatedly, we can obtain the data of each pair of $\{x_{\mu_{1}}, y_{\mu_{1}}\}\cup\{s,s'\}$ in $G_{\mu_{1}}$ by computing
$$F_1 (\mu_k ) \oplus \left (F_3 (\{\mu_{1},\mu_{2}\} ) \oplus \cdots \oplus F_3 (\{\mu_{k-1},\mu_{k}\} ) \right ) = F_1 (\mu_k ) \oplus \left ( \oplus_{i=1}^{k-1} F_3 (\{\mu_{i},\mu_{i+1}\} ) \right ).$$
Similarly, by computing
$$ (F_3 (\{\lambda_{1},\lambda_{2}\} ) \oplus \cdots \oplus F_3 (\{\lambda_{\ell-1},\lambda_{\ell}\} )  ) \oplus F_1 (\lambda_\ell ) = \left ( \oplus_{j=1}^{\ell-1} F_3 (\{\lambda_{j},\lambda_{j+1}\} ) \right ) \oplus F_1 (\lambda_\ell ),$$
we can obtain the data of each pair of $\{x_{\lambda_{1}}, y_{\lambda_{1}}\}\cup\{t,t'\}$ in $G_{\lambda_{1}}$.

Furthermore, $F_4 (\{\mu_1,\lambda_1\} )$ contains the data of each pair of $\{x_{\mu_1}, y_{\mu_1}\}\cup\{x_{\lambda_1}, y_{\lambda_1}\}$ in $G\setminus E (G_{\mu_1} )\setminus E (G_{\lambda_1} )$.
Therefore, by combining this and the results of the above two operations, we can obtain the data of each pair of $\{s,s'\}\cup\{t,t'\}$ in $G$. And the combined result can be expressed as
$$K_{\theta,\theta'}^{\vec{w}_{s,t}} = F_1 (\mu_k ) \oplus \left ( \oplus_{i=1}^{k-1} F_3 (\{\mu_{i},\mu_{i+1}\} ) \right )
\oplus F_4 (\{\mu_1,\lambda_1\} ) \oplus
\left ( \oplus_{j=1}^{\ell-1} F_3 (\{\lambda_{j},\lambda_{j+1}\} ) \right ) \oplus F_1 (\lambda_\ell ).
$$
Then, we obtain the distance and the beer distance by $\kdis{G}{s}{t} = w_{s,t} (s,t ), \kbdis{G}{s}{t} = w_{s,t}^{\mathrm{B}} (s,t )$.

Here, the computation of $K_{\theta,\theta'}^{\vec{w}_{s,t}}$ can be written
$$K_{\theta,\theta'}^{\vec{w}_{s,t}} = F_1 (\mu_k ) \oplus \left ( \widehat{\oplus}_{i=1}^{k-1} F_3 (\{\mu_{i},\mu_{i+1}\} ) \right )
\oplus F_4 (\{\mu_1,\lambda_1\} ) \oplus
\left ( \widehat{\oplus}_{j=1}^{\ell-1} F_3 (\{\lambda_{j},\lambda_{j+1}\} ) \right ) \oplus F_1 (\lambda_\ell )
$$
by using the semigroup $\widehat{\oplus}$. Therefore, if we preprocess $\kct$ for Tree Product Query regarding $\widehat{\oplus}$, we can compute $K_{\theta,\theta'}^{\vec{w}_{s,t}}$ for the $\oplus,\widehat{\oplus}$ operation in $O (\alpha (|V (\kct )| ) )=O (\alpha (m ) )$ time.

\section{Algorithm based on tree decomposition}\label{sec:treedecomposition}
In this section, we describe algorithms based on tree decomposition.
For a graph $G$ with $n$ vertices and $m$ edges, denote its treewidth by $t\coloneqq \mathrm{tw}(G)$. Also, let $\kct$ be a rooted tree decomposition of $G$ with width $t$ and number of nodes $O(tn)$.
The following theorem holds for the space of data structures to be constructed, the preprocessing time, and the query time.
\begin{theorem}
    \begin{enumerate}
        \item When we construct a data structure in $O(t^3n)$ space with preprocessing using $O(t^8n)$ time, we can answer a query in $O(t^8n+\alpha(tn))$ time.
        \item When we construct a data structure in $O(t^3n)$ space with preprocessing using $O(t^8n)$ time, we can answer a query in $O(t^7+\alpha(tn))$ time.
        \item When we construct a data structure in $O(t^5n)$ space with preprocessing using $O(t^{10}n)$ time, we can answer a query in $O(t^6+\alpha(tn))$ time.
    \end{enumerate}
\end{theorem}

\medskip 

In the following, we describe an outline of the proof of the above theorems.

For each node $\mu$ in $\kct$, let $X_\mu$ be the vertex subset of $G$ that $\mu$ has, and let $S_\mu = X_\mu \cup \bigcup_{\lambda \in \kdes{\mu}}X_\lambda$.
Furthermore, let $A_\mu$ be a vertex in $X_\mu$ if $\mu$ is a root node, and if $\mu$ is not the root node, $A_\mu = X_\mu \cap X_\lambda$ where $\mu \in \kch{\lambda}$ ($A_\mu$ corresponds to the endpoint set of $\krefe{\mu}$ in the SPQR tree).

Then, we define the following symbols as well as the mapping to the SPQR tree.
\begin{align*}
    \kdsVec{1}{\mu}{u}{v} &\coloneqq \vec{d}(G[S_\mu],u,v)\ (\mu \in V(\kct), u,v\in A_\mu),\\
    \kdsVec{2}{\mu}{u}{v} &\coloneqq \vec{d}(G\setminus E(G[S_\mu]),u,v) \ (\mu \in V(\kct), u,v\in A_\mu),\\
    \kdsVec{3}{\{\mu,\lambda\}}{u}{v} &\coloneqq \vec{d}(G[S_\mu]\setminus E(G[S_\lambda]),u,v) \ ((\mu,\lambda) \in E(\kct), \lambda\in \kch{\mu}, u,v\in A_\mu\cup A_\lambda),\\
    \kdsVec{4}{\{\lambda,\lambda'\}}{u}{v} &\coloneqq \vec{d}(G\setminus E(G[S_\lambda])\setminus E(G[S_{\lambda'}]),u,v) \\
    &\ (\{\lambda,\lambda'\} \in \binom{\kch{\mu}}{2}, \mu\in V(\kct), u,v\in A_\lambda\cup A_\lambda').
\end{align*}
We can calculate $\vec{f}_1$ as follows: If $\mu$ is a leaf of $\kct$, then $\kdsD{1}{\mu}{u}{v}=\kdis{G[X_\mu]}{u}{v}$ and
$$\kdsBD{1}{\mu}{u}{v}=\left\{\begin{matrix}
    \min_{p\in B\cap X_\mu} \{\kdis{G[X_\mu]}{u}{p}+\kdis{G[X_\mu]}{p}{v}\} & B\cap X_\mu \neq  \emptyset \\
    \infty & B\cap X_\mu =  \emptyset.
\end{matrix}\right.$$
If $\mu$ is not leaves, then
$$\kdsD{1}{\mu}{u}{v}=\min \left\{\begin{matrix}
    \kdis{G[X_\mu]}{u}{v}\\
    \min_{\substack{\lambda\in\kch{\mu}\\p,q\in A_\lambda}} \{\kdis{G[X_\mu]}{u}{p}+\kdsD{1}{\lambda}{p}{q}+\kdis{G[X_\mu]}{q}{v}\}
\end{matrix}\right\},$$
$$\kdsBD{1}{\mu}{u}{v}=\left\{\begin{matrix}
    \min\left\{ \ell_{u,v},\min_{p\in B\cap X_\mu} \{\kdis{G[X_\mu]}{u}{p}+\kdis{G[X_\mu]}{p}{v}\} \right\} & B\cap X_\mu \neq  \emptyset \\
    \ell_{u,v} & B\cap X_\mu =  \emptyset.
\end{matrix}\right.$$
where
$$\ell_{u,v}=\min_{\substack{\lambda\in\kch{\mu}\\p,q\in A_\lambda}} \{\kdis{G[X_\mu]}{u}{p}+\kdsBD{1}{\lambda}{p}{q}+\kdis{G[X_\mu]}{q}{v}\}.$$
$\vec{f}_2,\vec{f}_3,\vec{f}_4$ can be calculated using the same idea.

For each $\mu \in V(\kct)$, $|X_\mu|=O(t),|E(G[X_\mu])|=O(t^2)$, so $\kAlgD{G[X_\mu]}=O(t^2)$ is obtained regardless of how we take the range $W$ of the weights.
Noting this, the space $C_i^{\mathrm{space}}$ and computation time $C_i^{\mathrm{time}}$ required for each $\vec{f}_i$ can be evaluated as follows.

$$C_1^{\mathrm{space}},C_2^{\mathrm{space}}=\sum_{\mu\in V(\kct)}O\left(|A_\mu|^2\right)=O(t^2 |V(\kct)|)=O(t^3n),$$
$$C_3^{\mathrm{space}}=\sum_{\kce\in E(\kct)}O(t^2)=O(t^2 |E(\kct)|)=O(t^3n),$$
$$C_4^{\mathrm{space}}=\sum_{\mu\in V(\kct)}\sum_{\psi\in \binom{\kch{\mu}}{2}} O(t^2)=O(t^4 |V(\kct)|)=O(t^5n),$$
$$C_1^{\mathrm{time}},C_2^{\mathrm{time}}=\sum_{\mu\in V(\kct)} \sum_{u,v\in A_\mu} \sum_{\lambda\in\kch{\mu}} \sum_{p,q\in A_\lambda}O\left( \kAlgD{G[X_\mu]}\right)=O(t^7 |V(\kct)|)=O(t^8n),$$
$$C_3^{\mathrm{time}}=\sum_{(\mu,\lambda)\in E(\kct)} \sum_{u,v\in A_\mu\cup A_\lambda} \sum_{\theta\in\kch{\mu}\setminus\{\lambda\}} \sum_{p,q\in A_\theta} O(t^2)=O(t^7 |E(\kct)|)=O(t^8n),$$
$$C_4^{\mathrm{time}}=\sum_{\mu\in V(\kct)}\sum_{\{\lambda,\lambda'\}\in\binom{\kch{\mu}}{2}} \sum_{u,v\in A_\lambda\cup A_{\lambda'}} \sum_{\theta\in\kch{\lambda,\lambda'}\setminus\{\lambda\}} \sum_{p,q\in A_\theta} O(t^2)=O(t^9 |V(\kct)|)=O(t^{10}n).$$
Then, we consider the computational complexity of queries when these are precomputed.
First, if $\vec{f}_1,\vec{f}_2$ are precomputed, the query can be solved in $O(t^7|V(\kct)|+\alpha(|V(\kct)|)+t^6)=O(t^8n+\alpha(tn))$ time. Next, if $\vec{f}_1,\vec{f}_2,\vec{f}_3$ are precomputed, the query can be solved in $O(t^7+\alpha(|V(\kct)|)+t^6)=O(t^7+\alpha(tn))$ time. Finally, if $\vec{f}_1,\vec{f}_2,\vec{f}_3,\vec{f}_4$ are precomputed, the query can be solved in $O(\alpha(tn)+t^6)$ time.

Here, the $t^6$ term appearing in each computational time is the time required to perform $O(1)$ times operation ($\oplus$) to integrate the data structure without using Tree Product Query (for $\widehat{\oplus}$). Of course, the $t^6$ term could be replaced by $\alpha(tn)$ if a better semigroup could be defined.

These computation complexity shows that the degree of $t$ in each result is larger than that of $r$ in the case of triconnected component decomposition (SPQR tree). The dominant factor of this is that $u,v$ in each $\kdsVec{i}{\cdot}{u}{v}$ can be taken in $O(t^2)$ ways in the tree decomposition, whereas $O(1)$ ways in triconnected component decomposition.

\section{Figures and Tables}

\begin{figure}[h]
    \centering
    \includegraphics[keepaspectratio, scale=0.5]{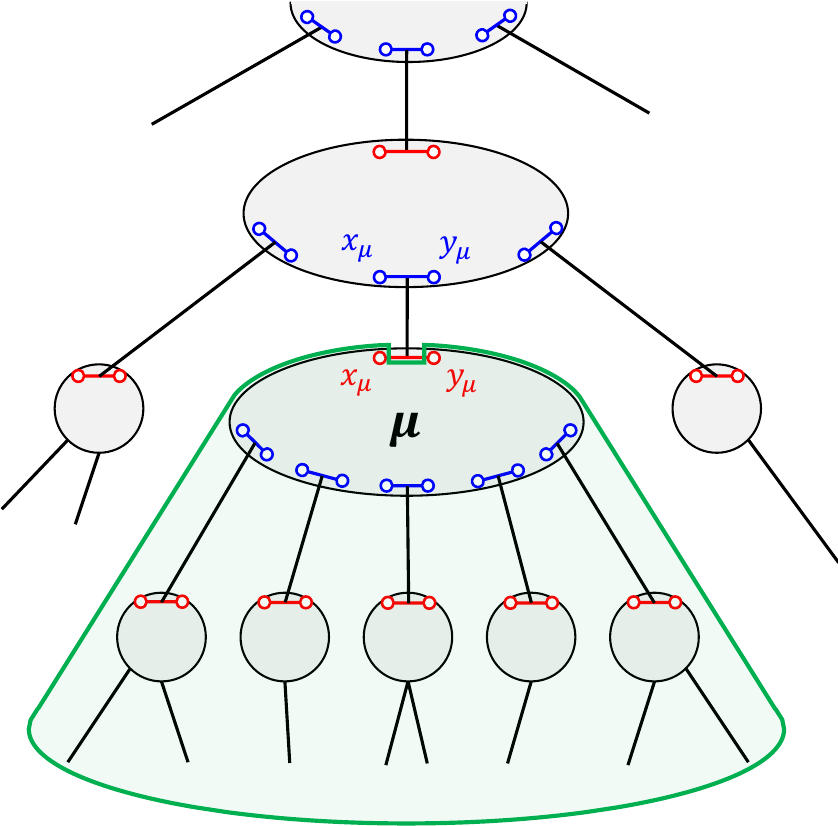}
    \caption{The subtree considered in $F_1(\mu)$}
    \label{fig:SubgraphForF1}
\end{figure}

\begin{table}[h]
  \caption{The weights $\kdsVec{1}{\mu}{u}{v}$ in the case $\mu$ is a Q node}
  \label{table:F1_Qnode}
  \centering
  \begin{tabular}{c||c}
  \hline
    & $\kdsVec{1}{\mu}{u}{v}$ \\
   \hline \hline
   $B\cap \{x_{\mu},y_{\mu}\}=\emptyset$ & $\begin{pmatrix}w (u,v )\\ \infty \end{pmatrix}$ \\
   \hline
   $B\cap \{x_{\mu},y_{\mu}\}\neq \emptyset$ & $\begin{pmatrix}w (u,v )\\ \displaystyle \min_{p\in B\cap \{x_{\mu},y_{\mu}\}} \{w (u,p )+w (p,v )\} \end{pmatrix}$  \\
   \hline
  \end{tabular}
\end{table}

\begin{figure}[h]
    \centering
    \includegraphics[keepaspectratio, scale=0.45]{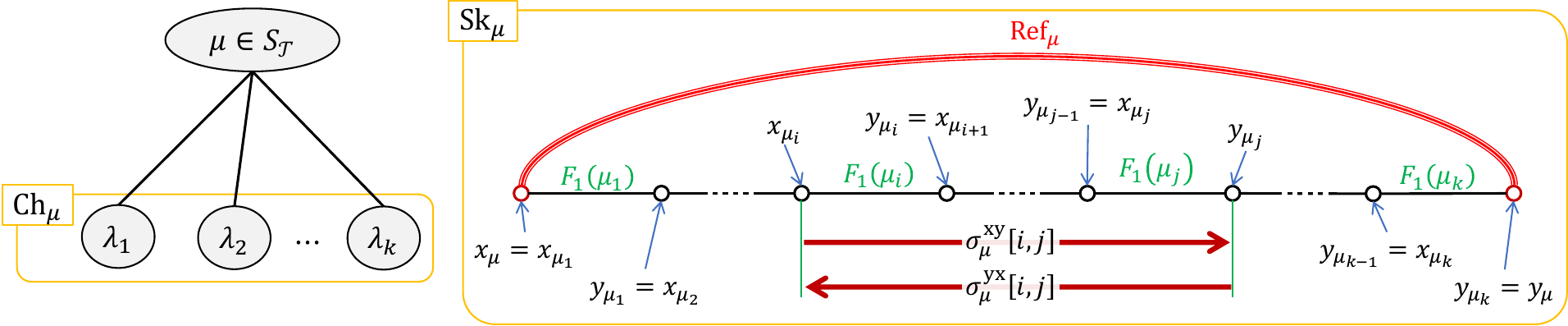}
    \caption{Notation for skeleton of S node}
    \label{fig:SnodeAndChildren}
\end{figure}

\begin{table}[h]
\caption{The weights $\kdsVec{1}{\mu}{u}{v}$ in the case $\mu$ is a S node}
  \label{table:F1_Snode}
  \centering
\begin{tabular}{c||c}
    \hline
    $\langle u,v \rangle$ & $\kdsVec{1}{\mu}{u}{v}$ \\
    \hline \hline
    $\langle x_{\mu},x_{\mu} \rangle$ & $\begin{pmatrix}0\\ \min_{1\leq i \leq k} \left\{\kbxx{\mu}{i} \right\} \end{pmatrix}$\\
    \hline
    $\langle x_{\mu},y_{\mu} \rangle$ & $\begin{pmatrix} \ksxy{\mu}{1,k} \\ \ksxy{\mu}{1,k} + \min_{1\leq i \leq k} \left\{ \kbxy{\mu}{i} \right\}\end{pmatrix}$ \\
    \hline
    $\langle y_{\mu},x_{\mu} \rangle$ & $\begin{pmatrix} \ksyx{\mu}{1,k} \\ \ksyx{\mu}{1,k} + \min_{1\leq i \leq k} \left\{ \kbyx{\mu}{i} \right\}\end{pmatrix}$ \\
    \hline
    $\langle y_{\mu},y_{\mu} \rangle$ & $\begin{pmatrix}0\\ \min_{1\leq i \leq k} \left\{\kbyy{\mu}{i} \right\} \end{pmatrix}$ \\
    \hline
\end{tabular}
\end{table}

\begin{table}[h]
\caption{The weights $\kdsVec{1}{\mu}{u}{v}$ in the case $\mu$ is a P node}
  \label{table:F1_Pnode}
  \centering
\begin{tabular}{c||c}
    \hline
    $\langle u,v \rangle$ & $\kdsVec{1}{\mu}{u}{v}$ \\
    \hline \hline
    $\langle x_{\mu},x_{\mu} \rangle$ & $\begin{pmatrix}0\\ \min \left\{ \ell^\mathrm{B}_{x_\mu,x_\mu} ,\  \ell_{x_\mu,y_\mu}+\ell^\mathrm{B}_{y_\mu,y_\mu}+\ell_{y_\mu,x_\mu},\  \ell^\mathrm{B}_{x_\mu,y_\mu}+\ell_{y_\mu,x_\mu},\  \ell_{x_\mu,y_\mu}+\ell^\mathrm{B}_{y_\mu,x_\mu} \right\} \end{pmatrix}$\\
    \hline
    $\langle x_{\mu},y_{\mu} \rangle$ & $\begin{pmatrix} \ell_{x_\mu,y_\mu} \\ \min \left\{ \ell^\mathrm{B}_{x_\mu,x_\mu}+\ell_{x_\mu,y_\mu},\ \ell_{x_\mu,y_\mu}+\ell^\mathrm{B}_{y_\mu,y_\mu},\ \ell^\mathrm{B}_{x_\mu,y_\mu},\ 2\ell_{x_\mu,y_\mu}+\ell^\mathrm{B}_{y_\mu,x_\mu} \right\}\end{pmatrix}$ \\
    \hline
    $\langle y_{\mu},x_{\mu} \rangle$ & $\begin{pmatrix} \ell_{y_\mu,x_\mu} \\ \min \left\{ \ell^\mathrm{B}_{y_\mu,y_\mu}+\ell_{y_\mu,x_\mu},\ \ell_{y_\mu,x_\mu}+\ell^\mathrm{B}_{x_\mu,x_\mu},\ \ell^\mathrm{B}_{y_\mu,x_\mu},\ 2\ell_{y_\mu,x_\mu}+\ell^\mathrm{B}_{x_\mu,y_\mu} \right\}\end{pmatrix}$ \\
    \hline
    $\langle y_{\mu},y_{\mu} \rangle$ & $\begin{pmatrix}0\\ \min \left\{ \ell^\mathrm{B}_{y_\mu,y_\mu} ,\  \ell_{y_\mu,x_\mu}+\ell^\mathrm{B}_{x_\mu,x_\mu}+\ell_{x_\mu,y_\mu},\  \ell^\mathrm{B}_{y_\mu,x_\mu}+\ell_{x_\mu,y_\mu},\  \ell_{y_\mu,x_\mu}+\ell^\mathrm{B}_{x_\mu,y_\mu} \right\} \end{pmatrix}$ \\
    \hline
\end{tabular}
\end{table}

\begin{figure}[h]
    \centering
    \includegraphics[keepaspectratio, scale=0.55]{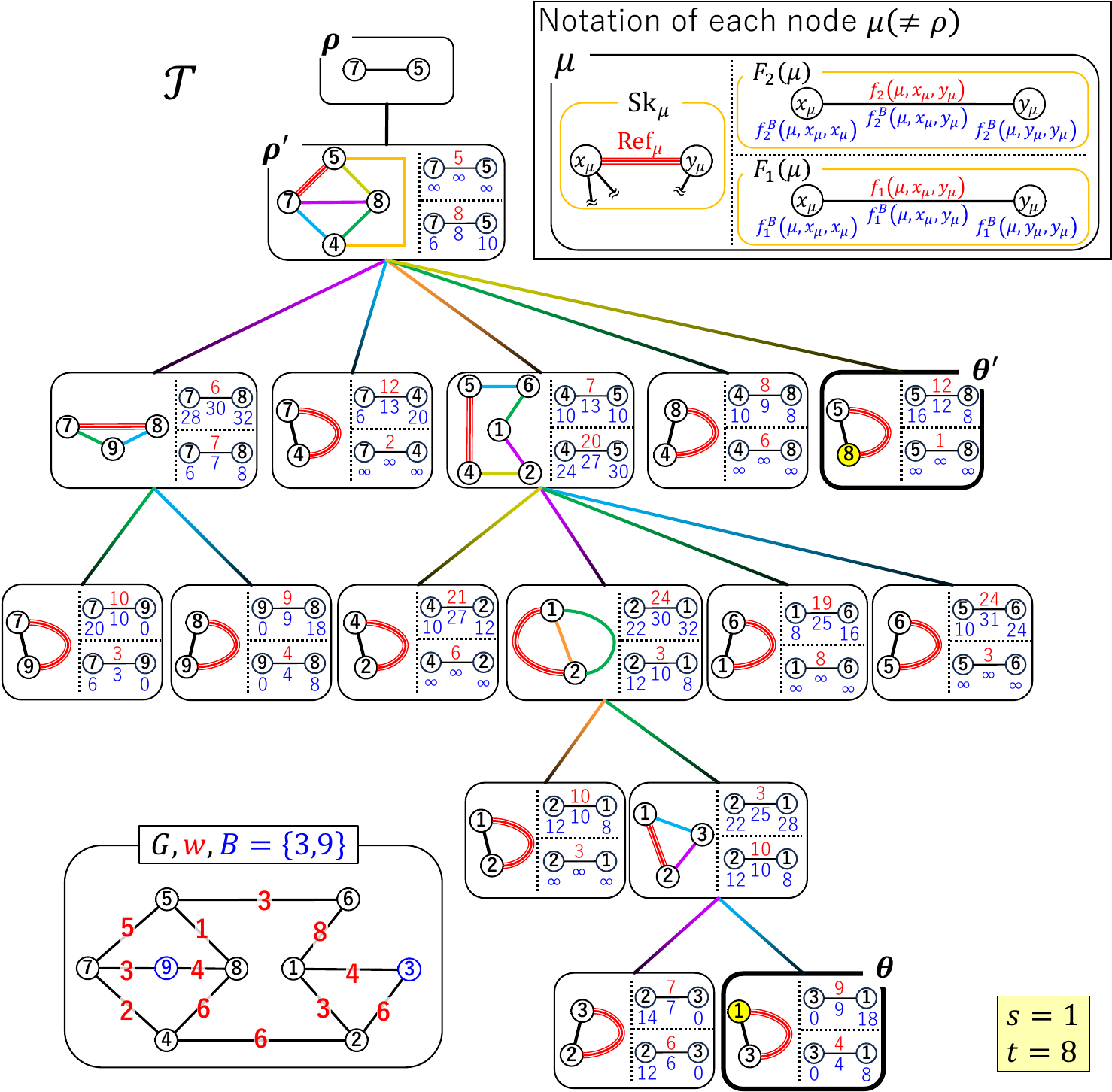}
    \caption{An weighted beer graph $(G,w,B)$ (lower left) and its SPQR tree with $F_1,F_2$}
    \label{fig:SPQRtree_F1F2}
\end{figure}

\begin{table}[h]
\caption{The weights $\kdsVec{3}{\kce}{u}{v}$ in the case $\mu$ is a S node and $u,v \in \{x_{\mu}, x_{\mu_i}\}, i\geq 2$}
  \label{table:F2_Snode}
  \centering
\begin{tabular}{c||c}
    \hline
    $\langle u,v \rangle$ & $\kdsVec{3}{\kce}{u}{v}$ \\
    \hline \hline
    $\langle x_{\mu},x_{\mu} \rangle$ & $\begin{pmatrix}0\\ \min_{1\leq j \leq i-1} \left\{ \kbxx{\mu}{j} \right\} \end{pmatrix}$\\
    \hline
    $\langle x_{\mu},x_{\mu_i} \rangle$ & $\begin{pmatrix} \ksxy{\mu}{1,i-1} \\ \ksxy{\mu}{1,i-1} + \min_{1\leq j \leq i-1} \left\{ \kbxy{\mu}{j} \right\}\end{pmatrix}$ \\
    \hline
    $\langle x_{\mu_i},x_{\mu} \rangle$ & $\begin{pmatrix} \ksyx{\mu}{1,i-1} \\ \ksyx{\mu}{1,i-1} + \min_{1\leq j \leq i-1} \left\{ \kbyx{\mu}{j} \right\}\end{pmatrix}$ \\
    \hline
    $\langle x_{\mu_i},x_{\mu_i} \rangle$ & $\begin{pmatrix}0\\ \min_{1\leq j \leq i-1} \left\{ \kbyy{\mu}{j} \right\} -(\ksyx{\mu}{i,k}+\ksxy{\mu}{i,k}) \end{pmatrix}$ \\
    \hline
\end{tabular}
\end{table}

\newpage

\end{document}